\documentclass[12pt]{article}
\usepackage{a4}
\usepackage{epsfig}
\usepackage{axodraw}

\evensidemargin=\oddsidemargin
\begin{document}

\newcommand{\kint}{\int \!\! \frac{dk}{(2 \pi)^d}}
\newcommand{\cfun}{C_0}
\newcommand{\ccfun}{C_2}
\newcommand{\dfun}{D_0}
\newcommand{\ddfun}{D_2}
\newcommand{\be}{\begin{equation}}
\newcommand{\ee}{\end{equation}}
\newcommand{\bea}{\begin{eqnarray}}
\newcommand{\eea}{\end{eqnarray}}
\newcommand{\beas}{\begin{eqnarray*}}
\newcommand{\eeas}{\end{eqnarray*}}
\newcommand{\cL}{{\cal L}}
\newcommand{\cR}{{\cal R}}
\newcommand{\cV}{{\cal V}}
\newcommand{\cH}{{\cal H}}
\newcommand{\cO}{{\cal O}}
\newcommand{\cM}{{\cal M}}
\newcommand{\td}{{\tilde d}}
\newcommand{\tu}{{\tilde u}}
\newcommand{\tg}{{\tilde g}}
\newcommand{\tn}{{\tilde \nu}}
\newcommand{\tl}{{\tilde l}}
\newcommand{\tC}{{\tilde C}}
\newcommand{\kslash}{k \!\!\! / }
\newcommand{\pslash}{p \!\!\! / }
\newcommand{\ra}{\rightarrow}

\newcommand{\MULL}[1]{{(M^2_U)_{LL}^{#1}}}
\newcommand{\MULR}[1]{{(M^2_U)_{LR}^{#1}}}
\newcommand{\MURL}[1]{{(M^2_U)_{RL}^{#1}}}
\newcommand{\MURR}[1]{{(M^2_U)_{RR}^{#1}}}

\newcommand{\Zp}{{\tilde Z_+}}

\renewcommand{\theequation}{\thesection.\arabic{equation}}

\newcommand{\vcb}{|V_{cb}|}
\newcommand{\vtd}{|V_{td}|}
\newcommand{\vub}{|V_{ub}/V_{cb}|}
\newcommand{\vts}{|V_{ts}|}

\def\eps{\varepsilon}
\def\epe{\varepsilon'/\varepsilon}
\newcommand{\gev}{\, {\rm GeV}}
\newcommand{\mev}{\, {\rm MeV}}
\newcommand{\Lms}{\Lambda_{\overline{\rm MS}}}

\newcommand{\mt}{m_{\rm t}}
\newcommand{\mtb}{\overline{m}_{\rm t}}
\newcommand{\mcb}{\overline{m}_{\rm c}}
\newcommand{\mc}{m_{\rm c}}
\newcommand{\ms}{m_{\rm s}}
\newcommand{\md}{m_{\rm d}}
\newcommand{\mb}{m_{\rm b}}
\newcommand{\mw}{M_{\rm W}}

\def\as{\alpha_s}

\newcommand{\imlt}{\IM\lambda_t}
\newcommand{\relt}{\RE\lambda_t}
\newcommand{\relc}{\RE\lambda_c}

\newcommand{\bi}{\begin{itemize}}
\newcommand{\ei}{\end{itemize}}
\newcommand{\ord}{{\cal O}}

\newcommand{\RE}{{\rm Re}}
\newcommand{\IM}{{\rm Im}}

\def\kpn{K^+\rightarrow\pi^+\nu\bar\nu}

\def\klpn{K_{\rm L}\rightarrow\pi^0\nu\bar\nu}

\def\kspn{K_{\rm S}\rightarrow\pi^0\nu\bar\nu}

\newcommand{\Ctilde}{\tilde{C}}

\newcommand{\kmm}{K_{\rm L} \to \mu^+ \mu^-}

\thispagestyle{empty}

{\normalsize\sf
\rightline {hep-ph/0408142}
\rightline{TUM-HEP-551/04}
\rightline{PITHA-04/13}
\vskip 3mm
\rm\rightline{August 10, 2004}
}

\vskip 5mm

\begin{center}
  
  {\LARGE\bf \boldmath{$\kpn$} and \boldmath{$\klpn$}  Decays \\[2mm]

 in the General MSSM}

\vskip 10mm

{\large\bf Andrzej J.~Buras$^1$, Thorsten Ewerth$^1$, Sebastian
J\"ager$^{1,2}$,}\\[1mm]
{\large\bf and Janusz Rosiek$^{1,3}$} \\[5mm]

{\small $^1$ Physik Department, Technische Universit{\"a}t M{\"u}nchen,}\\
{\small D-85748 Garching, Germany}\\
{\small $^2$ Institut f{\"u}r Theoretische Physik E, RWTH Aachen}\\
{\small D-52056 Aachen, Germany}\\
{\small $^3$ Institute of Theoretical Physics, Warsaw University}\\
{\small Ho\.za 69, 00-681 Warsaw, Poland}

\end{center}
\vskip 5mm

\renewcommand{\baselinestretch}{1.3}

\begin{abstract}
We reanalyze the rare decays $\kpn$ and $\klpn$ in a general MSSM with
conserved R-parity. Working in the mass eigenstate basis and
performing adaptive scanning of a large space of supersymmetric
parameters, 16 parameters in the constrained scan and 63 in the extended
scan, we find that large departures from the Standard Model
expectations are possible while satisfying all existing
constraints. Both branching ratios can be as large as a few times
$10^{-10}$ with $Br(\klpn)$ often larger than $Br(\kpn)$ and close to
its model independent upper bound. We give examples of supersymmetric
parameters for which  large departures from the SM expectations can
be found  and emphasize that the present $90\%$ C.L. experimental upper
bound on $Br(\kpn)$ gives a non trivial constraint on the MSSM parameter
space. Unlike previous analyses, we find that chargino box diagrams
can give, already for moderately light charged sleptons, a significant
contribution. As a byproduct we find that the ranges for the angles 
$\beta$ and $\gamma$ in the unitarity triangle are relaxed due to the
presence of new CP-violating phases in $K^0 -\bar K^0$ and
$B^0_d-\bar B^0_d$ mixing to 
$12^\circ \le \beta\le 27^\circ$ and $20^\circ\le\gamma\le 110^\circ$.
\end{abstract}

\newpage 
\setcounter{page}{1} 
\setcounter{footnote}{0}

\section{Introduction}
\label{sec:intro}
\setcounter{equation}{0}

The rare decays $\kpn$ and $\klpn$ belong to the theoretically
cleanest processes in the field of meson decays.  In fact their
branching ratios can be computed to an exceptionally high degree of
precision that is not matched by any other decay of
mesons~\cite{BB1,BB2,MU98,BB98,BB3}.  A detailed review of
$K\to\pi\nu\bar\nu$ decays within the Standard Model (SM) and its
extensions has been recently presented in~\cite{BSU1}, where an
extensive list of relevant papers can be found.  Earlier reviews have
been given in~\cite{BBL,Gino03}.

As emphasized in~\cite{BSU1}, the clean theoretical character of these
decays remains valid in essentially all extensions of the SM, whereas
this is generally not the case for non-leptonic two-body B decays used
to determine the CKM parameters through CP asymmetries and/or other
strategies.  While several mixing induced CP asymmetries in
non-leptonic B decays within the SM are essentially free from hadronic
uncertainties~\cite{FLREV}, as the latter cancel out due to the
dominance of a single CKM amplitude, this is often not the case in
extensions of the SM in which the amplitudes receive new contributions
with different weak phases implying no cancellation of hadronic
uncertainties in the relevant observables~\cite{CUSI02,Neu04}.

In this context an important virtue of $K\to\pi\nu\bar\nu$ decays is
the possibility of parametrizing the new physics contributions to
their branching ratios, in essentially all extensions of the SM, in a
model-independent manner by just two parameters~\cite{BRS}, the
magnitude of the short distance function $X$ and its complex phase:
\be\label{NX}
X=|X|e^{i\theta_X}
\ee
with $|X|=X(x_t)$ and $\theta_X=0$ in the SM.  
Here $x_t=m_t^2/M_W^2$.

The most recent predictions for the relevant branching ratios within
the SM read~\cite{BSU1}
\begin{equation}\label{SMkp0}
Br(\kpn)_{\rm SM}=(7.8 \pm 1.2)\cdot 10^{-11},
\ee
\be\label{SMkl0}
 Br(\klpn)_{\rm SM}= (3.0 \pm 0.6)\cdot 10^{-11},
\end{equation}
in the ballpark of other
estimates~\cite{Gino03,D'Ambrosio:2001zh,kettel,BFRS-II,BFRS-III}.
As discussed
in~\cite{BSU1} a NNLO calculation of the charm contribution to $\kpn$
and further progress on the determination of the CKM parameters coming
in the next few years dominantly from BaBar, Belle, Tevatron and later
from LHC and BTeV, should eventually allow predictions for
$Br(\kpn)$ and $Br(\klpn)$ with uncertainties of at most $\pm
5\%$.

On the experimental side the two events observed by the AGS E787
collaboration at Brookhaven~\cite{Adler970,Adler02} and the additional
event observed by AGS E949~\cite{E949} imply
\be\label{EXP1}
Br(\kpn)=(14.7^{+13.0}_{-8.9})\cdot 10^{-11} .
\ee
While the central value in~(\ref{EXP1}) is about a factor of two
higher than the SM value, the large experimental error precludes any
claims for signals of new physics in the $\kpn$ data.  Further
progress is expected in principle from AGS E949, from the efforts at
Fermilab around the CKM experiment~\cite{CKMEXP}, the corresponding
efforts at CERN around the NA48 collaboration~\cite{NA48EXP} and at
JPARC in Japan~\cite{JPAR}.  Hopefully this will bring additional
50-100 events in the next five years.

On the other hand the present experimental upper bound on $Br(\klpn)$
from KTeV~\cite{KTeV00X} reads
\be\label{EXP2}
Br(\klpn)<5.9 \times 10^{-7}.
\end{equation}
This is about four orders of magnitude above the SM expectation but a
$\klpn$ experiment at KEK, E391a~\cite{E391}, which recently took
data, should in its first stage improve the bound in~(\ref{EXP2}) by
three orders of magnitude.  While this is insufficient to reach the SM
level, a few events could be observed if $Br(\klpn)$ turned out to be
by one order of magnitude larger due to new physics contributions.
Further progress on this decay is expected from KOPIO~\cite{KOPIO} at
Brookhaven, and from the second stage of the E391 experiment at
JPARC~\cite{JPAR}, that could in principle provide a few hundreds of
$\klpn$ SM events, which would be truly fantastic!

In this context let us recall that a model-independent upper bound on
$Br(\klpn)$ following from isopin symmetry reads~\cite{GRNIR}
\begin{equation}
\label{iso}
Br(\klpn) < 4.4 \cdot Br(\kpn).
\end{equation}
With the data in~(\ref{EXP1}), which imply \cite{E949} 
\be
Br(\kpn) < 3.8 \cdot 10^{-10}~(90\%~\mbox{C.L.}),
\ee
one finds then~(\ref{iso})
\be\label{absbound}
Br(\klpn) < 1.7 \cdot 10^{-9}~(90\%~\mbox{C.L.}),
\ee
still two orders of magnitude below the upper bound from the KTeV
experiment in~(\ref{EXP2}).

Recently it has been pointed out in~\cite{BFRS-II,BFRS-III} that the
anomalies seen in the $B\to\pi K$ data may imply a significantly
enhanced value of $|X|$ and a large complex phase $\theta_X$:
\be\label{rX}
|X|=2.17\pm 0.12, \qquad  \theta_X= -(86\pm 12)^\circ,  
\ee
to be compared with $X=1.53\pm 0.04$ and $\theta_X=0$ in the SM.

The implications of this possibility are spectacular:
\begin{equation}\label{NPkp}
Br(\kpn)=(7.5 \pm 2.1)\cdot 10^{-11},
\ee
\be\label{NPkl}
 Br(\klpn)= (3.1 \pm 1.0)\cdot 10^{-10}.
\end{equation}
Consequently, while $Br(\klpn)\approx (1/3)Br(\kpn)$ in the SM, it is
substantially larger than $Br(\kpn)$ in this scenario.  The huge
enhancement of $Br(\klpn)$ found in~\cite{BFRS-II,BFRS-III} is mainly
due to the large weak phase $\theta_X$ and its negative sign.
One also finds
\be
\frac{{Br}(\klpn)}{{Br}(\kpn)}
\approx 4.2\pm 0.2~,
\ee
which means that $Br(\klpn)$ is rather close to its model-independent
upper bound in~(\ref{iso}).

A spectacular implication of these findings is a strong violation of
the golden minimal flavour violation  relation~\cite{BB4}
\be\label{R7}
(\sin 2\beta)_{\pi\nu\bar\nu}=(\sin 2\beta)_{\psi K_S},
\ee
where $-\beta$ is the phase of the $V_{td}$ coupling that could in
general be violated by new physics contributions to $K\ra \pi\nu\bar
\nu$ and the asymmetry in $B_d\ra\psi K_S$.

Indeed, one finds~\cite{BFRS-II,BFRS-III}
\be
(\sin 2 \beta)_{\pi \nu\bar\nu}=\sin 2(\beta- \theta_X)
=-(0.69^{+0.23}_{-0.41}),
\ee
in striking disagreement with $(\sin 2 \beta)_{\psi K_{\rm S}}=
0.736\pm0.049$.  Other interesting implications are discussed
in~\cite{BFRS-II,BFRS-III,India,Gino04}.

The analysis in~\cite{BFRS-II,BFRS-III} was model independent within a
scenario in which new physics manifested itself only through enhanced
EW penguins with a new complex phase.  This scenario was first
considered in~\cite{BRS,COLISI,Buras:1998ed,Buras:1999da} in the
context of rare $K$ decays and the ratio $\epe$ measuring direct CP
violation in the neutral kaon system, and was generalized to rare $B$
decays in~\cite{BHI}.  In~\cite{BFRS-II,BFRS-III} it has been further
generalized to $B\to\pi K$ decays.

The question that we would like to address in the present paper is
whether the spectacular effects in $K\to\pi\nu\bar\nu$ system found
in~\cite{BFRS-II,BFRS-III} are still possible within the general MSSM
with new sources of flavour violation when all known experimental and
theoretical constraints on the relevant parameters are taken into
account.

New flavour violating interactions are present in the general MSSM
because the sfermion mass matrices can be non-diagonal in the basis in
which all quark-squark-neutral-gaugino vertices and quark and lepton
mass matrices are flavour diagonal.
Instead of diagonalizing sfermion mass
matrices it is often convenient to consider their off-diagonal terms
as new flavour violating interactions.  This so-called mass insertion
approximation (MIA)~\cite{HKR} has been reviewed in the classic
papers~\cite{GGMS,MPR,GRNIRA}, where further references can be found.

In the context of the $K\to\pi\nu\bar\nu$ decays the most extensive
analyses using the mass insertion method can be found
in~\cite{NIRWOR,BRS,COLISI,Buras:1999da}.  It turns out, that sizeable
enhancements of $K\to\pi\nu\bar\nu$ rates can only be generated by
chargino-mediated diagrams with a large LR mixing in the up-squark sector.
Reference~\cite{Buras:1999da}, the most detailed of these
papers, finds the upper bounds
\be\label{BSNEW}
Br(\kpn)\le 1.7 \cdot 10^{-10}, \qquad Br(\klpn)\le 1.2 \cdot 10^{-10}.
\ee
Larger values were found rather unlikely.  Moreover, as discussed in
detail in~\cite{BRS}, in these models the minimal flavour violation
 relation in~(\ref{R7})
can be violated due to the presence of a new phase $\theta_X$.  A
rough estimate shows that this phase could be as large as $\pm
25^\circ$.  This is not as large as found in~\cite{BFRS-II,BFRS-III}
but sizable departures from the SM as seen in~(\ref{BSNEW}) can be
found.

However, in~\cite{BRS,Buras:1999da} the following assumptions
(approximations) have been made:

\begin{itemize}
\item 
MIA (up to the second order of expansion) 
was used in calculation of $K\ra \pi\nu\bar\nu$ decay rates and other
related processes.
\item limited number of MSSM parameters, assumed to be most important, 
were used in numerical scans.
\item bounds on  mass insertions coming from various experimental 
results were obtained by the approximate procedure of requiring that
each individual term in the mass insertion expansion at most saturates the
measured value; this neglects the possibility of  significant
cancellations between contributions from different  terms in
the expansion and also 
interference with the SM contribution that is always present.
\item some additional assumptions not coming from low-energy
physics, like GUT induced RGE relations between mass insertions, were
taken into account.
\end{itemize}

The questions that we would like to address here are then, whether by
relaxing the assumptions made in~\cite{BRS,Buras:1999da}, going beyond
the MIA and performing more extensive
numerical scans, while satisfying all existing experimental
constraints on supersymmetric parameters,
\begin{itemize}
\item
$\theta_X$ can be as large as found in~\cite{BFRS-II,BFRS-III},
\item
$Br(\kpn)$ can be significantly enhanced over the SM expectations so
that it is at least as high as its central experimental value
in~(\ref{EXP1}),
\item
$Br(\klpn)$ can be enhanced by an order of magnitude with the ratio
$Br(\klpn)/Br(\kpn)$ reaching the bound in~(\ref{iso}).
\end{itemize}

Moreover, there are the questions:
\begin{itemize}
\item
In what regions of the parameter space of the general MSSM do the
maximal departures from the SM expectations occur?
\item
What are the bounds on the MSSM parameters implied already by the
experimental value on $Br(\kpn)$ in~(\ref{EXP1})?
\end{itemize}

Answering these questions is a non-trivial numerical task, due to the
large number of free parameters and experimental constraints which
have to be considered.  Here we would like to demonstrate an efficient
method of a random scan over the MSSM parameter space, based on an
adaptation of the Monte Carlo integration algorithm
VEGAS~\cite{BREIN,VEGAS}.  Such a method is designed to
automatically concentrate most of the randomly generated points in the
SUSY parameter ranges giving the largest deviations from the SM
results, thus allowing for analyzing very large parameter spaces, with
20 or more dimensions, in a reasonable time and without very extensive
computer CPU usage. This allows us to answer the first three of
the questions listed above. The remaining two points will be discussed
elsewhere.

Our paper is organized as follows.  In Section~\ref{sec:basic} we will
present the formulae for the branching ratios $Br(\kpn)$ and
$Br(\klpn)$ in terms of the function $X$, discuss their properties and
compare the general MSSM considered here with the simple new physics
scenario analyzed in \cite{BFRS-II,BFRS-III}.  In
Section~\ref{sec:num} we discuss the problematic of the huge space of
parameters in supersymmetric theories and the method that we propose
to tackle this problem.  In Section~\ref{sec:results} we present our
numerical results.  Finally, in Section~\ref{sec:conclusions} we
summarize the main findings of our paper.
 A list of the contributions to the function $X$ from the
relevant box and penguin diagrams is given in an~\ref{app:wilson}.

\section{Basic Formulae}
\label{sec:basic}
\setcounter{equation}{0}
\subsection{Effective Hamiltonian}
The effective Hamiltonian relevant for $\kpn$ and $\klpn$ decays in a
general MSSM considered in our paper can be written as follows
\begin{equation}\label{hMSSM} 
{\cal H}_{\rm eff}={G_{\rm F} \over{\sqrt 2}}{\alpha\over 2\pi 
\sin^2\theta_{\rm w}}
\left[ 
{\cal H}_{\rm eff}^{(c)}+{\cal H}_{\rm eff}^{(t)}\right]
\end{equation}
where the internal charm part
\be\label{Hc}
{\cal H}_{\rm eff}^{(c)}=
\sum_{l=e,\mu,\tau} V^{\ast}_{cs}V_{cd} X^l_{\rm NL}
 (\bar sd)_{V-A}(\bar\nu_l\nu_l)_{V-A} \, 
\end{equation}
is fully dominated by the SM contributions and 
\be\label{Ht}
{\cal H}_{\rm eff}^{(t)}=
\sum_{l=e,\mu,\tau} V^{\ast}_{ts}V_{td} 
\left[X_L (\bar sd)_{V-A}(\bar\nu_l\nu_l)_{V-A}+ 
X_R (\bar sd)_{V+A}(\bar\nu_l\nu_l)_{V-A}\right]
\end{equation}
with $X_L$ receiving both the SM and supersymmetric contributions and
$X_R$ only the latter ones that also include charged Higgs
exchanges. In the SM $X_L$ is real and given for 
$\overline{m}_t(m_t)= (168.1\pm 4.1)~\gev$ by
\bea
X_L^{SM}\equiv X_{SM}= 1.53\pm 0.04~.
\label{eq:xsm}
\eea

The index $l$ = $e$, $\mu$, $\tau$ denotes the lepton flavour.  The
dependence on the charged lepton mass resulting from the box diagrams
is negligible for the top contribution.  In the charm sector this is
the case only for the electron and the muon but not for the
$\tau$-lepton.  Below we will give the branching ratios that follow
from~(\ref{Ht}).

This parametrization of ${\cal H}_{\rm eff}^{(t)}$ is very useful for
phenomenological applications and for the comparison
with~\cite{BFRS-II,BFRS-III}, but one should remember that in a
general MSSM scenario considered here not all contributions to ${\cal
H}_{\rm eff}^{(t)}$ are proportional to $V^{\ast}_{ts}V_{td}$.  This
means that parametrizing ${\cal H}_{\rm eff}^{(t)}$ as in~(\ref{Ht}),
necessarily puts some CKM dependence into $X_L$ and $X_R$.  Therefore
we will use below also a different parametrization of new physics
contributions.

Now, as the strong interactions are not sensitive to the sign of
$\gamma_5$, the hadronic matrix elements of $(\bar s d)_{V-A}$ and
$(\bar s d)_{V+A}$ are equal to each other, and consequently the
function $X$ playing the central role in our paper is simply given by
\be
X=X_L+X_R=|X| e^{i \theta_X}~.
\label{eq:x}
\ee
Explicit expressions for $X_L$ and $X_R$ are collected in
the~\ref{app:wilson}.  As one can see there, chargino and part of the
neutralino contributions to $X_R$ are strongly suppressed by small
Yukawa couplings of the down quarks.  As we will see, the remaining
neutralino contribution proportional to $U(1)$ gauge coupling is
typically also much smaller than the dominant terms in $X_L$, so that
$|X_R|\ll |X_L|$ also in the MSSM (except may be for some
non-interesting points where $X_L$ is small due to cancellations) and
can be neglected for all practical purposes.

\subsection{Branching Ratios}

The branching ratios for the $K\ra \pi \nu \bar \nu$ decays, in the
new physics scenario considered, can be written as follows \cite{BSU1}
\begin{equation}
\label{bkpnZ}
{Br}(\kpn)=\kappa_+\left[\left({{\IM}(\lambda_t X)\over
\lambda^5}\right)^2 + \left({{\RE}\lambda_c\over \lambda}P_c(X) 
+ {{\RE}(\lambda_t X)\over\lambda^5}\right)^2\right],
\end{equation}
\begin{equation}
\label{bklpnZ}
Br(\klpn)=\kappa_L \left({{\IM}(\lambda_t X)\over
\lambda^5}\right)^2
\end{equation}
with $\lambda=0.224$ being one of the Wolfenstein
parameters~\cite{WO}, $\lambda_t=V^{\star}_{ts}V_{td}$,
$\lambda_c=V^{\star}_{cs}V_{cd}$, $\kappa$-factors equal
to~\cite{BSU1}
\begin{equation}\label{kappi}
\kappa_+=(4.84\pm 0.06)\cdot 10^{-11}, \qquad
\kappa_L=(2.12\pm 0.03)\cdot 10^{-10}~
\end{equation}
and
\begin{equation}\label{p0k}
P_c(X)=\frac{1}{\lambda^4}\left[\frac{2}{3} X^e_{\rm NL}+\frac{1}{3}
 X^\tau_{\rm NL}\right]=0.39\pm 0.07
\end{equation}
resulting from the NLO calculations in~\cite{BB98,BB3}.  The anatomy
of the error in $P_c(X)$ has been recently presented in~\cite{BSU1}.

As discussed in the next section we will use as our input parameters
\be\label{INPUT}
|V_{us}|=\lambda=0.224, \qquad \vcb=A\lambda^2=0.0415, 
\qquad
R_b=0.37
\ee
where
\be
R_b=\frac{(1-\lambda^2/2)}{\lambda}\left|\frac{V_{ub}}{V_{cb}}\right|.
\ee

We recall that $A$ is a Wolfenstein parameter~\cite{WO} with
$A=0.83\pm0.02$ and $R_b$ is one of the sides of the unitarity
triangle.

While the parameters in~(\ref{INPUT}) contain uncertainties, the
latter are sufficiently small so that they can be neglected in
comparison with hudge uncertainties in the values of supersymmetric
parameters. The choice of these three parameters is dictated by the
fact that they are extracted from tree-level decays and consequently
their values are not subject to new physics uncertainties.

As the fourth variable we will take the angle $\gamma$ in the
unitarity triangle that is equal to the CKM phase $\delta_{CKM}$ in
the standard parametrization of this matrix.  The angle $\gamma$ can
be in principle measured without any new physics pollution in tree-level
$B$ decay strategies that will be only available at LHC and
BTeV~\cite{FLREV}.  In the general MSSM the value of $\gamma$ may
deviate from the one extracted from the usual analysis of the
unitarity triangle that uses SM expressions. For this reason we will
allow $\gamma$ to vary in the full range
\be
-180^\circ \le \gamma \le 180^\circ,
\ee
but as we will see in Section 4, only the range $20^\circ\le\gamma\le
110^\circ$ is allowed when all constraints are taken into account.

We define next $\overline{X}_{SUSY}$ through
\be\label{XSUSY}
\lambda_t X =\lambda_t X_{SM}+\lambda^5 \overline{X}_{SUSY}
\ee 
and introduce 
\be
a=P_c(X)+\frac{\vcb^2}{\lambda^4} X_{SM}\approx 1.43, \qquad 
b=R_b \frac{\vcb^2}{\lambda^4} X_{SM}\approx 0.39,
\ee
that do not depend on supersymmetric parameters. We find then
\be\label{BKP1}
Br(\kpn)=\kappa_+\left[(Q_{eff}(\gamma)+\IM \overline{X}_{SUSY})^2 +
(P_{eff}(\gamma)-\RE \overline{X}_{SUSY})^2\right] ,
\ee
\be\label{BKP2}
Br(\klpn)=\kappa_L\left(Q_{eff}(\gamma)+\IM
\overline{X}_{SUSY}\right)^2~ ,
\ee
where
\be
P_{eff}(\gamma)=(1-\frac{\lambda^2}{2})\left[ a - b \cos\gamma
\right],\qquad Q_{eff}(\gamma)= b \sin\gamma~.
\ee
As $P_{eff}(\gamma)$ and $Q_{eff}(\gamma)$ can be fully determined
within the SM, provided $\gamma$ can be measured through tree-level
decays, the formulae (\ref{BKP1}) and (\ref{BKP2}) transparently
exhibit supersymmetric contributions. With $\gamma\approx 65^\circ$ we
have
\be
P_{eff}(\gamma)\approx 1.23,\qquad
Q_{eff}(\gamma)\approx 0.35~.
\ee

\subsection{Comparison with \cite{BFRS-III}}

In order to compare our results with the ones in \cite{BFRS-III}, it
is useful to have the expressions for both branching ratios in terms
of $R_t$ and $\beta$, where $\beta$ is another angle and $R_t$ another
side of the unitarity triangle.

Now, for each pair $(R_b,\gamma)$, used above, one can determine the pair $(R_t,\beta)$
through
\be\label{VTDG}
R_t=\sqrt{1+R_b^2-2 R_b\cos\gamma},\qquad
\cot\beta=\frac{1-R_b\cos\gamma}{R_b\sin\gamma}
\ee
and consequently calculate both branching ratios by using
\begin{equation}
\label{bkpnZ1}
{Br}(\kpn)=\kappa_+\left[\tilde r^2 A^4 R_t^2 |X|^2 +2 \tilde r \bar
P_c(X) A^2 R_t |X|\cos\beta_X + \bar P_c(X)^2 \right]
\end{equation}
\begin{equation}
\label{bklpnZ1}
Br(\klpn)=\kappa_L
\tilde r^2 A^4 R_t^2 |X|^2\sin^2\beta_X,
\end{equation}
where
\be\label{Pbar}
\bar P_c(X)=\left(1-\frac{\lambda^2}{2}\right) P_c(X),
\ee
\be\label{BX}
\beta_X=\beta-\beta_s-\theta_X, \qquad 
\tilde r=\left|\frac{V_{ts}}{V_{cb}}\right|
\approx 0.985,
\ee
\be\label{VTDVTS}
V_{td}=A R_t\lambda^3 e^{-i\beta},\qquad V_{ts}=-|V_{ts}|e^{-i\beta_s}.
\ee
The small phase $\beta_s\approx -1^\circ$ can be neglected for all
practical purposes.

While the formulae~(\ref{bkpnZ1}) and~(\ref{bklpnZ1}) look exactly as
the ones given in ~\cite{BFRS-III}, there is one important difference
that should be empasized here. In the new physics scenario considered
in~\cite{BFRS-III} the values of $\beta$, $R_t$ and $\gamma$ were unaffected
by new physics contributions and the sole effects of new physics were
felt only through $|X|$ and $\theta_X$.  In a general MSSM the
$a_{\psi K_S}$ asymmetry does not generally measure $\beta$ but rather
$\beta+\theta_d$ with $\theta_d$ coming from new complex contributions
in $B_d^0-\bar B^0_d$ mixing. Thus the constraint on the angle $\beta$
coming from the $a_{\psi K_S}$ asymmetry reads
\be\label{APSI}
\sin2(\beta+\theta_d)=0.736 \pm 0.049
\ee
so that $\beta$ could be quite different from $\beta=23.5^\circ$ found
in the SM and used in \cite{BFRS-III}. Moreover $R_t$ can be modified. 
Therefore, the comparison of the results obtained here with the ones in 
\cite{BFRS-III} requires some care.

\section{Numerical analysis}
\label{sec:num}
\setcounter{equation}{0}

\subsection{Independent MSSM parameters}
\label{subsec:par}

In the general MSSM, the predictions for the branching ratios considered
here and various experimental constraints can depend on almost every MSSM
parameter.
The dependence on the majority of them is weak enough, however,
to be safely neglected in principle for the purpose of computing
$Br(K_L\ra \pi^0 \nu \bar \nu)$ and
$Br(K^+\ra \pi^+ \nu \bar \nu)$.

Still, parameters may indirectly enter through the expressions for other
observables, which we use to constrain the parameter space.
Taking a conservative approach we allow for the independent variation of
more free parameters than included in the existing literature.

\begin{table}[htbp]
\begin{center}
\begin{tabular}{p{10cm}p{10mm}r}
Kobayashi-Maskawa phase &&
$-180^{\circ}\leq \gamma \leq 180^{\circ}$ \\
CP-odd Higgs boson mass && $150\leq M_A \leq 400$ \\
$SU(2)$ gaugino mass; we use $M_1$ GUT-related to $M_2$ &&
$50\leq M_2 \leq 800$ \\
Gluino mass && $195\leq m_{\tilde g} \leq 2000$ \\
Supersymmetric Higgs mixing parameter && $-400\leq \mu \leq 400$ \\
Common flavour diagonal slepton mass parameter &&
$95\leq M_{sl} \leq 1000$ \\
Common mass parameter for the first two generations && \\
of squarks, as well as $\tilde b_R$ && $240\leq M_{sq} \leq 1000$ \\
Squark mass parameter for the left stop and sbottom &&
$50\leq M_{\tilde t_L} \leq 1000$ \\
Squark mass parameter for the right stop &&
$50\leq M_{\tilde t_R}\leq 1000$ \\
Flavour universal trilinear scalar mixing parameter && \\
(normalized to the fermion mass) &&
$-1\leq A \leq 1$\\
Mass insertion $\delta_{LL}^{12}$ && $|\delta_{LL}^{12}|\leq 0.135$\\
Mass insertion $\delta_{ULR}^{13}$ && $|\delta_{ULR}^{13}|\leq 1.65$\\
Mass insertion $\delta_{ULR}^{23}$ &&
$|\delta_{ULR}^{23}|\leq 1.65$\\
\end{tabular}
\caption{Parameters and their ranges used in the ``constrained'' scan
(16 real degrees of freedom). All mass parameters are in GeV.
\label{tab:constr}}
\end{center}
\end{table}

To start with, we assume that $Br(K_L\ra \pi^0 \nu \bar \nu)$ and
$Br(K^+\ra \pi^+ \nu \bar \nu)$ may depend significantly on the set of
unknown SM and SUSY parameters listed in Table~\ref{tab:constr}.
We took as fixed the other SM parameters, including fermion masses and
the CKM parameters given in (\ref{INPUT}),
as their measurements are known to be relatively insensitive to physics beyond
the SM. This leaves as our only free SM parameter the
CKM angle $\gamma$, which to date has not been determined
from tree-level decays.

Apart from the flavour diagonal supersymmetric mass parameters, we vary several
off-diagonal ones. For a transparent parameterization and easy comparison
with the literature, it is useful to define mass insertions,
even though our actual computation makes use of exactly diagonalized
sfermion, chargino, and neutralino mass matrices.
Following~\cite{GGMS,MPR}, we define
\be
        \delta^{IJ}_{SXY} = \frac{(M^2_S)^{IJ}_{XY}}
                        {\sqrt{(M^2_S)^{II}_{XX}\,(M^2_S)^{JJ}_{YY}}} .
\ee
Here $I,J$ denote quark flavours, $X,Y$ the (superfield) chirality,
and the ``sector'' $S$ can be either
$U$ or $D$. Evidently, this definition depends on the basis
chosen in flavour space. For the right-handed up and down squarks,
there exists a preferred choice selected by the mass eigenstates of
their respective SM partners. In the left-handed sector this is different,
because $SU(2)$ invariance dictates that the MSSM Lagrangian contains only
one soft mass matrix for
the doublet squarks. This implies that $\delta^{IJ}_{ULL}$ and
$\delta^{IJ}_{DLL}$ are related. In the super-CKM basis, where each
squark field is the superpartner of a quark mass eigenstate, this
correspondence is given by CKM matrix elements.
As a consequence, even for diagonal
$(M^2_D)^{IJ}_{LL}$ the off-diagonal $\delta^{IJ}_{ULL}$ are in
general nonzero in this basis. Also, the mixings and mass splittings
among left-handed up squarks are constrained by bounds on down-squark
mixings from $K^0-\bar K^0$ mixing.

In the context of $K\ra \pi \nu \bar\nu$, it has been pointed out
by the authors of~\cite{BRS} that a particularly useful choice of basis
is one where the left-handed up squark basis states are simply
the $SU(2)$ partners of the left-handed down squarks, which in turn are the
superpartners of the left-handed down-type quarks, so that the gluino
and neutralino couplings are flavour conserving.
This eliminates the CKM elements from the chargino couplings to
the left-handed up squarks, and in particular
from the term in the MIA that is believed
to be dominant~\cite{COLISI}. Besides, such a basis ensures that any
choice of $\delta$'s automatically preserves $SU(2)$ invariance.
We use this basis, to be called the BRS basis, in presenting our results
in this paper. We also omit the ``sector'' index on the $\delta_{LL}$.

The decays $K_L\ra \pi^0 \nu \bar\nu$ and $K^+\ra \pi^+ \nu \bar\nu$ are
$\Delta S=1$ processes and 
correspond to a transition between the first and second quark generations.
Thus, they could certainly depend on $(12)$ squark mass insertions.
However, as pointed out in~\cite{COLISI}, both
decays are sensitive also to second-order terms in the mass insertion expansion, namely,
to products of a $(13)$ and (the conjugate of) a $(23)$ mass insertion. 
Therefore, we start
from varying independently five (12) mass insertions,
$\delta_{LL}^{12}$, $\delta_{DRR}^{12}$, $\delta_{URR}^{12}$,
$\delta_{DLR}^{12}$ and $\delta_{ULR}^{12}$, plus five more (13) and
five (23) mass insertions (assuming for the moment that all LR mass 
insertions are hermitian, e.g. $\delta_{DLR}^{13} =
\delta_{DLR}^{31\star}$).

\begin{figure}[htbp]
\begin{center}
\begin{tabular}{cc}
\epsfig{file=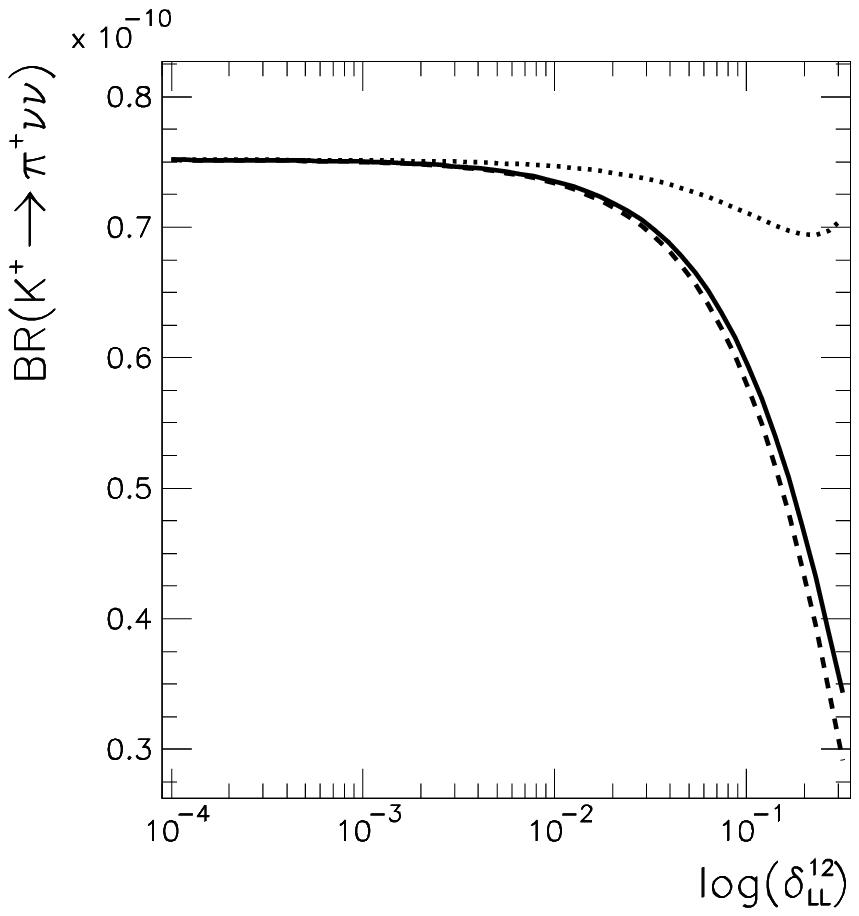,width=0.48\linewidth}
&
\epsfig{file=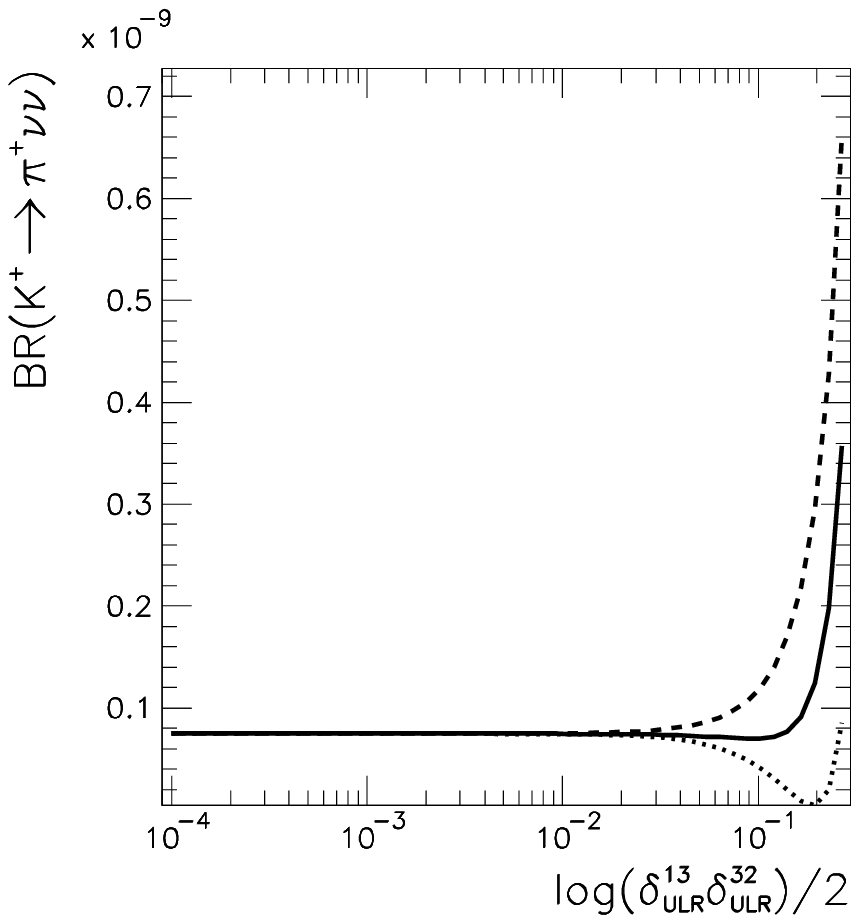,width=0.48\linewidth}
\\
\end{tabular}
\caption{Dependence of
$Br(K^+\ra \pi^+ \nu \bar\nu)$ on chosen mass insertions for
$\gamma=65^{\circ}$, $\tan\beta=2$, $M_{sq}=M_{\tilde{t}_L}=
M_{\tilde{t}_R}=500$,
$M_{sl}=300$, $M_2=200$, $m_{\tilde{g}}=3M_2$, $\mu=100$, $M_A=200$
(all masses in GeV). Solid, dashed and dotted lines: $\delta_{LL}^{12}$
and $\delta_{ULR}^{13}\delta_{ULR}^{32}$ complex, real and imaginary,
respectively.
\label{fig:sens_mi}}
\end{center}
\end{figure}

Even assuming that all flavour diagonal quantities from the list above
are real (there is no reason to constrain mass insertions to be real),
we have listed already 40 free real parameters.  This is a huge number
for any reasonably dense numerical scan.  To avoid excessive
computation time, we first tested how sensitive
to the various mass insertions the predictions for our branching ratios
really are.
In Fig.~\ref{fig:sens_mi} we plot the dependence of $Br(K^+\ra \pi^+
\nu \bar\nu)$ on $\delta_{LL}^{12}$ and the product $\delta_{ULR}^{13} 
\delta_{ULR}^{23\star}$ for a chosen set of SUSY parameters.  
The dependence on the other mass insertions, not shown in the plots, is
much weaker and can be neglected in the first approximation.

Fig.~\ref{fig:sens_mi} suggests that it is sufficient in the numerical
scan to vary
only three independent mass insertions, $\delta_{LL}^{12}$,
$\delta_{ULR}^{13}$ and $ \delta_{ULR}^{23}$. In fact the dependence
on  $\delta_{LL}^{12}$, which is tightly constrained from
mixing in the neutral kaon system, is almost negligible for realistic values.

Also, as we
tested, both branching ratios in question are only weakly sensitive to
GUT- or SUGRA-type assumptions concerning relations between
SUSY mass parameters (e.g unified gaugino masses or sfermion mass
parameters).  Thus, the large but already numerically more feasible
total number of 16 free parameters appears to be sufficient to explore the
possible ranges of $Br(K^+\ra \pi^+ \nu \bar \nu)$ and $Br(K_L\ra
\pi^0 \nu \bar \nu)$ in the framework of the MSSM.  However, this is not true
when trying to satisfy experimental bounds and selecting allowed values
of the SUSY parameters.  For example, keeping the product
$\delta_{LL}^{13} \delta_{LL}^{23\star}$ constant but varying
$\delta_{LL}^{13}$ independently does not affect $Br(K^+\ra \pi^+
\nu \bar\nu)$ and $Br(K_L\ra \pi^0 \nu \bar\nu)$ but allows to satisfy
bounds coming from $\Delta M_d$, the measured
$B^0_d-\bar B^0_d$ mass difference. This motivates keeping, in
$\delta_{LL}^{12}$, one more flavour violating parameter even in the
``minimal'' 16-parameter scan. Later, we will see that our numerical method
is powerful enough to study the consequences of further increasing
the number of degrees of freedom, which turn out to be unimportant.

\subsection{Theoretical and experimental bounds}
\label{subsec:bounds}

The enormous freedom in the SUSY parameter space is reduced by
a number of phenomenological constraints the theory must satisfy. From the
experimental side we take into account the set of bounds and measurements
listed in Table~\ref{tab:bounds}. As can be seen, we do not take into
account the $\varepsilon'/\varepsilon$ constraint, as its calculation still
has very large theoretical uncertainities. Doing this we are aware
of the fact that $\varepsilon'/\varepsilon$ could one day offer a very
powerful constraint on the size of SUSY contributions 
\cite{Buras:1998ed,Buras:1999da}.

\begin{table}[htbp]
\begin{center}
\begin{tabular}{lp{1mm}lp{1mm}l}
Quantity && Measured value && Experimental error \\ \hline
Lightest neutralino mass && $>$ 46.0 GeV && \\
Second lightest neutralino mass && $>$ 62.4 GeV && \\
Lightest chargino mass && $>$ 94.0 GeV && \\
The two ``sbottom'' masses && $>$ 89.0 GeV && \\
The two ``stop'' masses && $>$ 95.7 GeV && \\
all other squark masses && $>$ 250.0 GeV && \\
$|\varepsilon_K|$ && $2.280\cdot 10^{-3}$ && $0.013\cdot 10^{-3}$ \\
$\Delta M_K $ && $3.489\cdot 10^{-15}$ GeV && $0.008\cdot 10^{-15}$\\
$\Delta M_d $ && $3.31\cdot 10^{-13}$ GeV && $0.04\cdot 10^{-13}$ \\
$\Delta M_s $ && $> 9.5\cdot 10^{-12}$ GeV && \\
$Br(B_s\ra X_s \gamma)$ && $3.28 \cdot 10^{-4}$ &&
$\mbox{}^{+0.41}_{-0.36}\cdot 10^{-4}$\\
$(\sin 2 \beta)_{\psi K_{\rm S}}$ && $0.736$ &&$ 0.049$\\
\end{tabular}
\caption{Experimental measurements used to constrain the MSSM
parameter space. Limits on supersymmetric particles masses published
in~\cite{RPP2004} are used. The quotes on ``sbottom'' and ``stop''
indicate that these are only approximate flavour states.}
\label{tab:bounds}
\end{center}
\end{table}

Due to the large number of unknown parameters and substantial
theoretical uncertainities present in many calculations, it is
pointless (and certainly very difficult to implement from a practical
point of view) to use everywhere the
state-of-art calculations of higher order QCD corrections and other
non-leading effects calculated in the literature.
Nevertheless, whenever feasible we try to avoid
unnecessary simplifications made in many papers.  In particular:

\begin{itemize}
\item For low energy FCNC and CP violating processes,  we do not
restrict ourselves to the most commonly used MIA.
Instead, we calculate all Wilson coefficients relevant
for a given process in the mass eigenstate approach, taking into
account the full set of contributions - SM, Higgs, chargino, neutralino
and gluino sectors.  Further, we compare experimental results with the
full 1-loop SUSY expressions, not just the dominant (usually gluino)
term.
\item  To complete the bounds coming from the low-energy data, we also 
compare lower mass bounds for SUSY particles, obtained from
accelerator experiments, with the tree-level eigenvalues of the MSSM mass
matrices.
\end{itemize}

For comparing theoretical predictions for the low-energy
observables with experiment, we apply the following procedure.  For a
given set of SUSY parameters, we calculate all appropriate parton-level
diagrams.  Next, we construct expressions for the considered
quantities using the central values of the QCD evolution factors and
necessary hadronic matrix elements, obtained by perturbative SM and
lattice-QCD computations, respectively.  Finally, for every quantity $Q$
we require:
\bea
|Q^{exp} - Q^{th}| \leq 3\Delta Q^{exp} + q |Q^{th}|, \hskip 2cm q =
0.5
\label{eq:xacc}
\eea
with the exception of $\Delta M_s$, for which we
require $(1+q)|Q^{th}|\geq Q^{exp}$.

The first term on the RHS of (\ref{eq:xacc}) represents the $3\sigma$
experimental error.  The second term corresponds to the theoretical
error.  In principle, such an error differs from quantity to quantity
and is usually smaller than $q=50\%$, which we assumed as a generic
number in all calculations.  However, apart from the theoretical
errors coming from uncertainities in the QCD evolution and hadronic
matrix elements calculations, one should take into account also
problems arising due to the limited numerical scan density.  In principle,
with a very dense scan, it should be possible to find SUSY parameters
fulfilling~(\ref{eq:xacc}) within the ``true'' theoretical errors of
present calculations.  Such a dense scan requires, however, a huge amount
of computer time - with 16 or more free parameters and rather
complicated mass eigenstate formulae, it would take
months of CPU time.  This does not seem to be necessary and may even
be undesirable.  Our goal is to find ``generic'' allowed values for
the $K\ra\pi\nu\bar\nu$ decay rates, i.e. values possible to obtain for fairly
wide ranges of SUSY parameters, without strong fine-tuning and
resorting to some very particular points of the parameter space where
the experimental bounds are satisfied due to precise cancellations of
various types of contributions.  Thus, in our scan we use wide
``theoretical'' errors, assuming that this procedure points to the
correct ranges of the MSSM parameters, and if necessary the exact
values of such parameters fulfilling the bound  in (\ref{eq:xacc})
with smaller $q$ could always (or at least almost always) be
found. This could be achieved e.g. by a denser scan or by more advanced
numerical routines, possibly even solving numerically the set of
non-linear equations given by all constraints.

The only exception from the procedure described above is the
imposition of the constraint from the CP asymmetry $a_{\psi K_{\rm
S}}(t)$.  Making the safe assumption that the supersymmetric
contributions to the decay amplitude can be neglected, this asymmetry
measures the phase of the $B^0_d-\bar B^0_d$ amplitude
\bea
M_{12} = \langle B_d^0| H_{eff}(\Delta B=2)| \bar B^0_d \rangle =
|M_{12}| e^{i2(\beta+\theta_d)}
\eea
through
\bea
a_{\psi K_{\rm S}}(t) = - a_{\psi K_{\rm S}} \sin(\Delta M_d t) =
\sin2(\beta+\theta_d) \sin(\Delta M_d t)~.
\eea
Here $-\beta$ is the phase of $V_{td}$, while $\theta_d$ is the new
effective phase coming from the supersymmetric contributions.  As in the
general MSSM the presence of non-vanishing sfermion mass insertions
implies contributions of several local operators to $M_{12}$, the
phase $\theta_d$ suffers from potential uncertainities related to the
hadronic matrix elements of these operators. Only if one of these
operators dominates over the others in the full amplitude can
$\theta_d$ be cleanly related to supersymmetric parameters. In other
cases its error can be sometimes very large and moreover is hard to
estimate -- it would require additional scanning over hadronic
uncertainities for each given set of SUSY parameters.
To avoid such problems, in our analysis we assume conservatively 
\bea
|a_{\psi K_{\rm S}}^{exp} -\sin2(\beta+\theta_d)| \leq 2 \Delta
a_{\psi K_{\rm S}}^{exp}
\label{eq:psib}
\eea
with $\sin2(\beta+\theta_d)$ calculated in the MSSM for central
values of the hadronic parameters and $a_{\psi K_{\rm S}}^{exp} =
0.736\pm0.049$.  Expressing everything in degrees we require then
\bea
|23.7^{\circ}- (\beta+\theta_d)|\leq 4.2^{\circ} .
\label{eq:as_acc}
\eea
Assuming only experimental and no theoretical errors on the
R.H.S. of~(\ref{eq:psib}), we reject some otherwise valid points from
our scan, but we checked that this does not have any significant
effect on the results discussed in following sections.

Apart form the experimental bounds, there are also bounds from
the requirement that the vacuum is stable, or that the true ground state
of the theory does not break color and charge. We apply the corresponding
Charge and Color Breaking (CCB) and unbounded from below MSSM scalar
potential (UFB) bounds~\cite{CCBUFB}, which give constraints on the
$A$-terms and consequently on the left-right elements of the sfermion
mass matrices. We use the tree-level expressions in the form:
\bea
|A_u^{IJ}|^2 &\leq& (1 + 2\delta^{IJ}) f |Y_u^{max(I,J)}|^2
\left[(m^2_Q)^{II} + (m^2_U)^{JJ} + m_{H_2}^2 + |\mu|^2\right],\nonumber\\
|A_d^{IJ}|^2 &\leq& (1 + 2\delta^{IJ}) f |Y_d^{max(I,J)}|^2
\left[(m^2_Q)^{II} + (m^2_D)^{JJ} + m_{H_1}^2 + |\mu|^2\right],\nonumber\\
|A_l^{IJ}|^2 &\leq& (1 + 2\delta^{IJ}) f |Y_l^{max(I,J)}|^2
\left[(m^2_L)^{II} + (m^2_E)^{JJ} + m_{H_1}^2 + |\mu|^2\right],
\label{eq:ccb}
\eea
\bea
|A_u^{IJ}|^2 &\leq& f |Y_u^{max(I,J)}|^2 \left[(m^2_Q)^{II} + (m^2_U)^{JJ}
+ (m^2_L)^{KK} + (m^2_E)^{KK}\right],\nonumber\\
|A_d^{IJ}|^2 &\leq& f |Y_d^{max(I,J)}|^2 \left[(m^2_Q)^{II} + (m^2_D)^{JJ}
+ (m^2_L)^{KK}\right],\hskip 1cm I\neq J,\, K\neq I,J,\nonumber\\
|A_l^{IJ}|^2 &\leq& f |Y_l^{max(I,J)}|^2 \left[(m^2_L)^{II} + (m^2_E)^{JJ}
+ (m^2_L)^{KK}\right].
\label{eq:ufb}
\eea
As suggested in some papers (see e.g.~\cite{CCBLOOP} for a comprehensive
discussion), tree-level CCB/UFB bounds
may not be very stable against radiative corrections.  Again, we do not try
to use more elaborate formulae, but relax the tree-level bounds by
adding factors $f=2$ on the RHS of all equations in~(\ref{eq:ccb})
and~(\ref{eq:ufb}). (Strict tree-level bounds correspond to $f=1$.)

\subsection{The adaptive scan algorithm}
\label{subsec:scanadapt}

\begin{figure}[htbp]
\begin{center}
\begin{tabular}{cc}
\epsfig{file=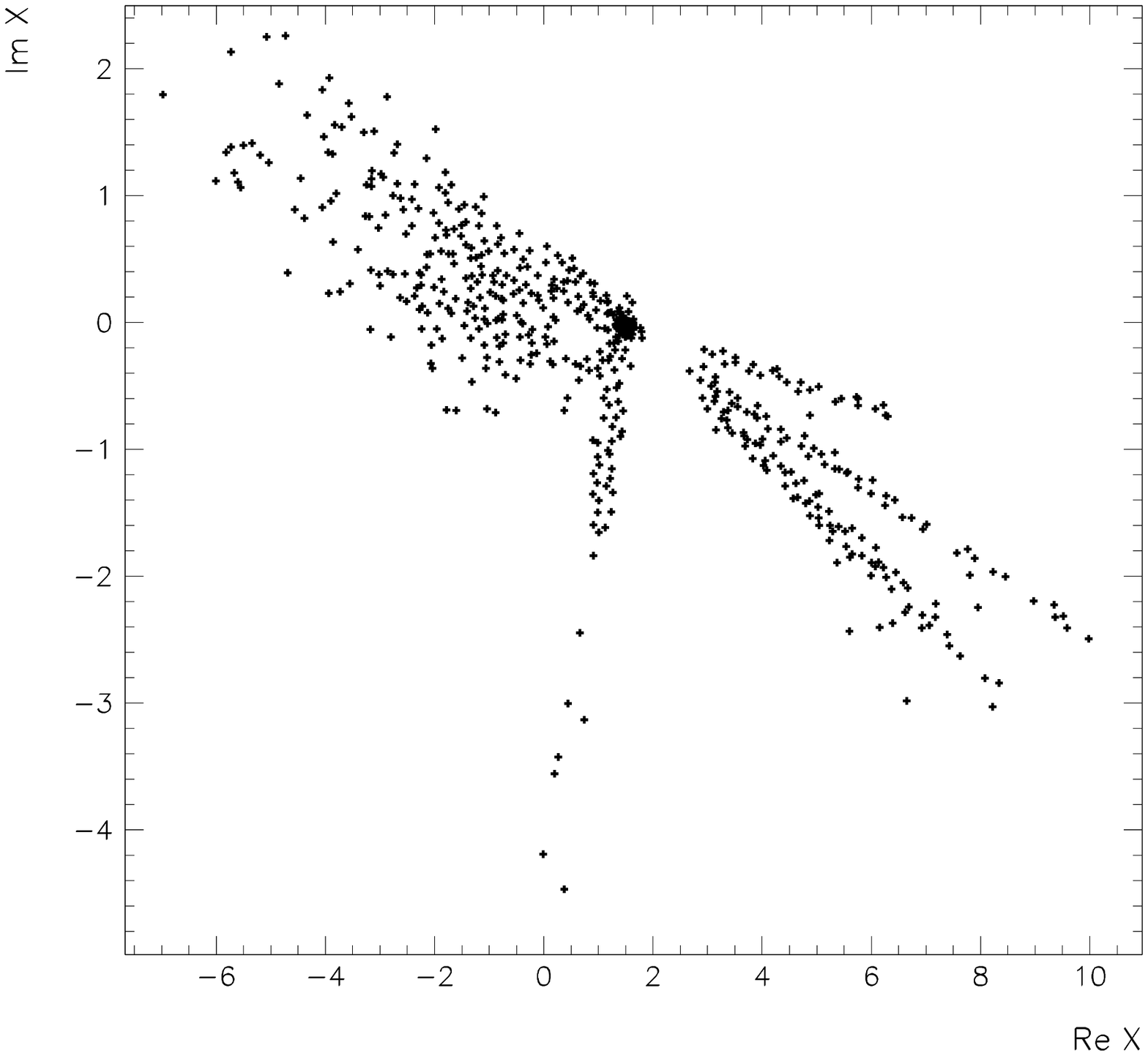,width=0.48\linewidth}
&
\epsfig{file=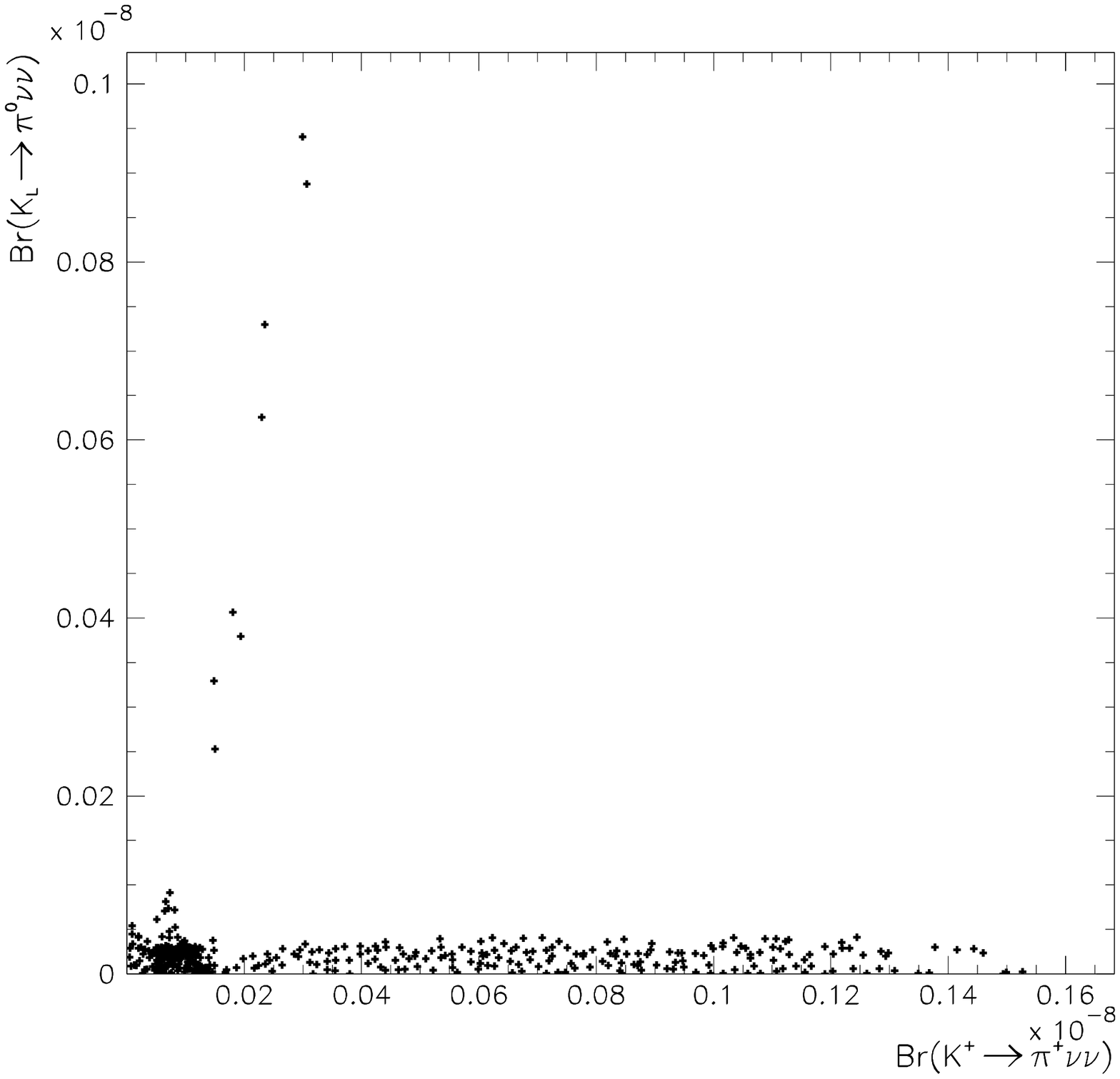,width=0.48\linewidth}
\\
\end{tabular}
\caption{Scatter plot of $X$,  $Br(K^+\ra \pi^+ \nu \bar \nu)$ 
and $Br(K_L\ra \pi^0 \nu \bar \nu)$ distributions for $\tan\beta=2$
and uniform grid scan.  Parameters varied in the ranges $0\leq
\gamma \leq 180^{\circ}$, $200\leq M_{sq}\leq 500$, $150\leq M_{sl} 
\leq 300$, $-400\leq \mu \leq 400$, $200\leq M_2 \leq 600$, $150\leq M_A \leq
300$, $-1\leq A \leq 1$, mass insertions between $0.003$ and $0.3$,
their phases set to $0^{\circ}$, $45^{\circ}$ or $90^{\circ}$.}
\label{fig:noadapt}
\end{center}
\end{figure}

The numerical analysis laid out in the previous subsections requires
us to scan an $N$-dimensional ($N \geq 16$) parameter space
sufficiently densely to account for all regions where the quantity $X$
of (\ref{eq:x}) possibly is large.  A straightforward approach is to
scan over a uniform $N$-dimensional grid.  However, this is both
time-consuming and ineffective.  Nevertheless, we have performed some
test scans of this kind to get a first notion of the attainable rates
of both branching ratios considered, as well as of $X$.  We generated
about 1.6M points, only about 1\% of which passed all theoretical and
experimental constraints.  The distribution of the remaining points is
shown in Fig.~\ref{fig:noadapt}.  What can be observed
is that most points are close to the SM value of $X$, i.e. new-physics
effects are very small for most grid points.  The departure from the
SM lies along straight lines, which is an obvious artifact of the
small grid density, in spite of the fact that more than 1 million
points were generated.

However, what we are interested in is not so much the somewhat vague
notion of a ``probability density'' for quantities shown in various
scatter plots, which is clearly a function of the choice of
parameters, of their scaling chosen during the scan (e.g.  linear or
logarithmic) and the range considered for each parameter.  Rather, we
are interested in the maximal allowed regions of the parameter space
and the extremal values of the observables, viz.  the (complex)
function $X$, which are physical (i.e. parameterization-invariant).
To find this information, a dense and very time-consuming scan is necessary.

In practice, due to the number of free parameters, the grid involves
only 2 to 5 different values for each parameter, posing an additional
risk that a relevant region of the parameter space may fall through
the grid altogether.  In fact, this will be likely if the largest
effects do not simply occur where some parameters are smallest and
others are largest, but in areas where some type of cancellation, even
moderate, is needed to fulfill the experimental constraints.

A random generation of parameter points may exhibit some of these
overlooked regions.  Such an approach is particularly useful when, as
it is true for the quantities considered in this paper, the analyzed
expressions depend strongly only on a subset of $M$ free parameters,
$M < N$.  In this case one can think in terms of an ``effective
density'' -- the projection of the random points onto the
$M$-dimensional subspace of ``relevant'' parameters now has a density
(in that subspace) that is about $(2\dots 5)^{N/M}$ times higher than for
a uniform grid. To see this, envision projecting the original grid:
for given relevant parameters, the remaining $N-M$ parameters
describing a given point, which are approximately irrelevant, just
parameterize points projected onto the same $M$-dimensional point.
Thus in the uniform-grid approach one repeats the same analysis many
times.  Conversely, the projection of the randomly distributed
points is again randomly distributed over the subspace, and each new
point gives additional information.

Note that in general the subset of relevant parameters is different
for each observable. This precludes just removing the irrelevant
parameters from the scan, in the case one believes to know them
beforehand.  The random scan, on the other hand, gives an improved
effective density for each of the corresponding subspaces.

Still, for randomly distributed parameter points, an expression
$|X-X^{SM}|$ describing the deviation from the SM prediction (or
similar expressions for other observables, which we use to constrain
the parameter space) will give a small number over most of the
parameter space. On the other hand, in the regions where it is large
it also tends to vary more strongly.  Therefore, while effectively
improving the grid density, random generation of points with a uniform
distribution may still miss the exact maxima for realistic numbers of
points.  It would be desirable to generate points more densely where
the function $X$ is large and/or varies quickly - such a procedure
would be particularly efficient if, as is expected, our results mainly
depend on just a few parameters.

Recently, Brein~\cite{BREIN} has suggested using an adaptive
Monte-Carlo integration algorithm such as VEGAS~\cite{VEGAS} for a
similar purpose.  VEGAS performs ``importance sampling'' via an
iterative algorithm.  At each iteration, it generates a certain number
of points according to a probability distribution determined by the
integrand values encountered during the previous iterations.  The
probability distribution is chosen to be separable and is adjusted
after each iteration.  The initial distribution is chosen to be
uniform.  This procedure is designed to minimize the statistical error
of the integration by increasing the number of points in those parts
of the integration volume where the integrand is found to be large.
While we are not interested in computing any integral to any given
precision, VEGAS does provide what we desire, if we choose our
integration volume as the parameter space (suitably rescaled) and the
integrand such that it becomes large for large values of $X$.  We
found the following ``integrand'' useful:
\bea
f = \left\{
\begin{array}{lp{1cm}l}
0 && \mathrm{parameter~set~rejected~by~constraints}\\
\left|X - X^{SM}\right|^n && \mathrm{constraints~satisfied}\\
\end{array}
\right.
\label{eq:vegfun}
\eea
where $X$ is the quantity defined in~(\ref{eq:x}), $X^{SM}$ is its SM
central value~(\ref{eq:xsm}) and we varied the power $n$ between 3 and
8. (We do not write explicitly the dependence of $X$ on $\gamma$
and SUSY parameters.) We ``integrated'' this function numerically with
the VEGAS routine, storing all generated Monte Carlo points in a
separate file along with the values of various observables for
analysis.  The parameter ranges used for the integration are given in
Table~\ref{tab:constr}.

Using the VEGAS algorithm has the great advantage of sampling
mostly the important regions of parameter space, where the function
$X$ really depends significantly on at least some of the parameters.
It also allows us to increase safely the number of the degrees of
freedom--- adding new variables has only a moderate effect on total
computation time as long as the function $X$, or in general the VEGAS
integrand, is weakly dependent on them.  In fact, we were able to
perform Monte Carlo sampling over huge portions of the MSSM parameter
space, probably never tried before, of up to 63 dimensions
(parameters), and to judge only afterwards, from the obtained
distributions, which of the parameters where important for a given
problem.

The applied procedure carries some dangers.  First, one should mention
a point that is not always appreciated.  As is well known, the
``random'' numbers provided by various standard libraries are in fact
deterministic sequences which may or may not satisfy certain
conditions of randomness, or absence of correlation.  In our context,
when one uses $N$
subsequent numbers in the sequence to define a point in an
$N$-dimensional parameter space,
the widely used generators based on linear congruences
tend to concentrate these points on a
bundle of $(N-1)$-dimensional hyperplanes.  This can reintroduce the
danger of points of interest ``falling through the grid'' mentioned
above.  To address this problem, we used two different random number
generators, both of which avoid such correlations.  The first is a
subtractive Mitchell-Moore generator discussed by Knuth~\cite{KNUTH}
in an implementation of Kleiss~\cite{KLEISS}.  The second is based on
a combination of Fibonacci and arithmetic sequences proposed
by~\cite{RANDOM2} in an implementation of  James~\cite{JAMES}.
We obtained
similar plots for both algorithms, which strengthens our conviction
that our procedure is stable and does not miss any relevant regions.

Also, there is the question of stability of data samples obtained in
this way.  In order to check it, we varied the power $n$
in~(\ref{eq:vegfun}) in the range of 3 to 8, and inspected how the
shape of the plotted regions evolved with the numbers of points
generated. The choice of the particular function $f$ appears to affect
mostly the time in which the shape of various distributions stabilizes,
not the boundaries itself. Similar comments apply to tuning of the
internal control parameters of the VEGAS routine.

\section{Results}
\label{sec:results}
\setcounter{equation}{0}

\subsection{Sensitivity to scan parameters}
\label{subsec:sens}

\begin{figure}[htbp]
\begin{center}
\begin{tabular}{cc}
\epsfig{file=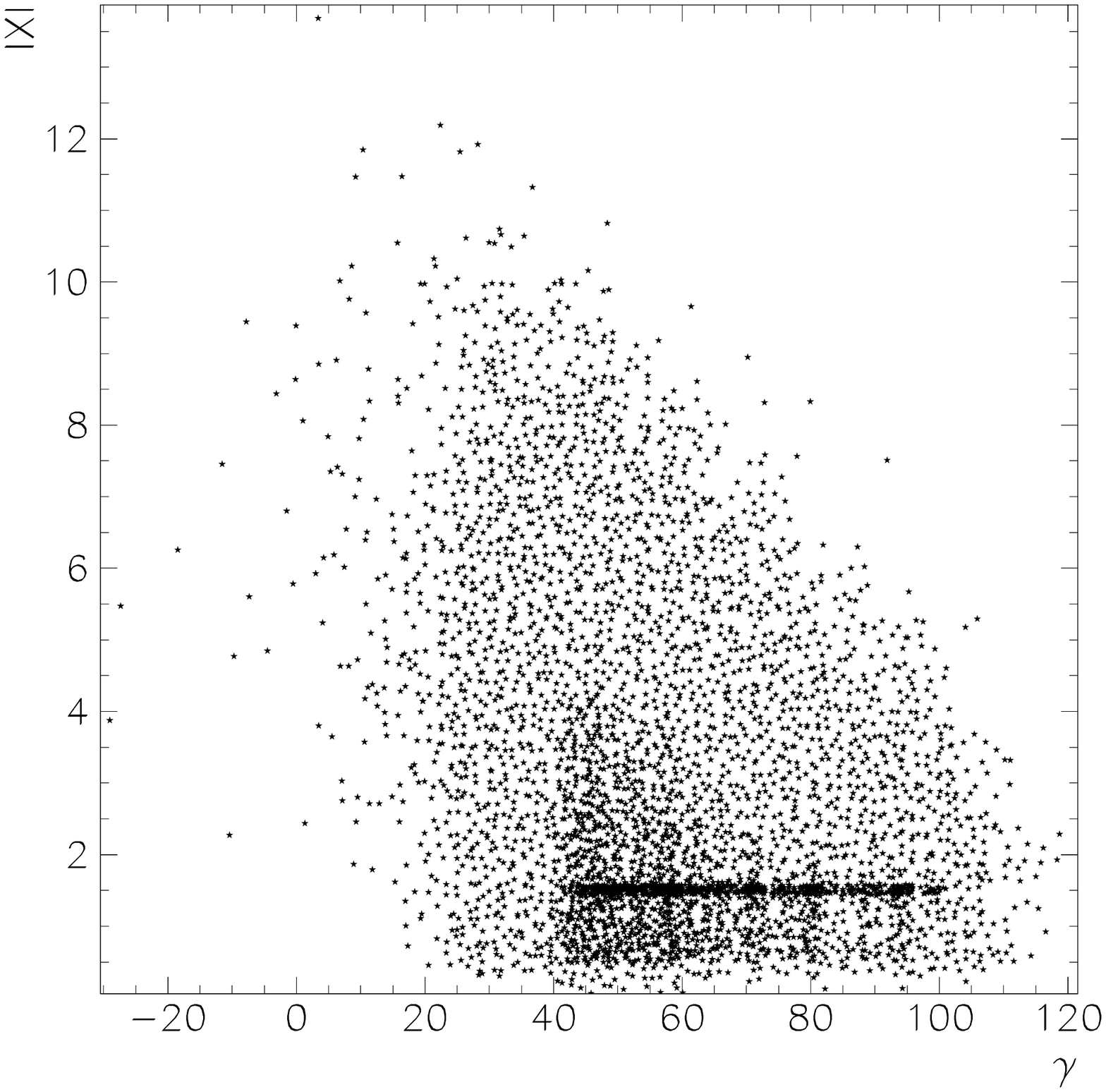,width=0.49\linewidth}
&
\epsfig{file=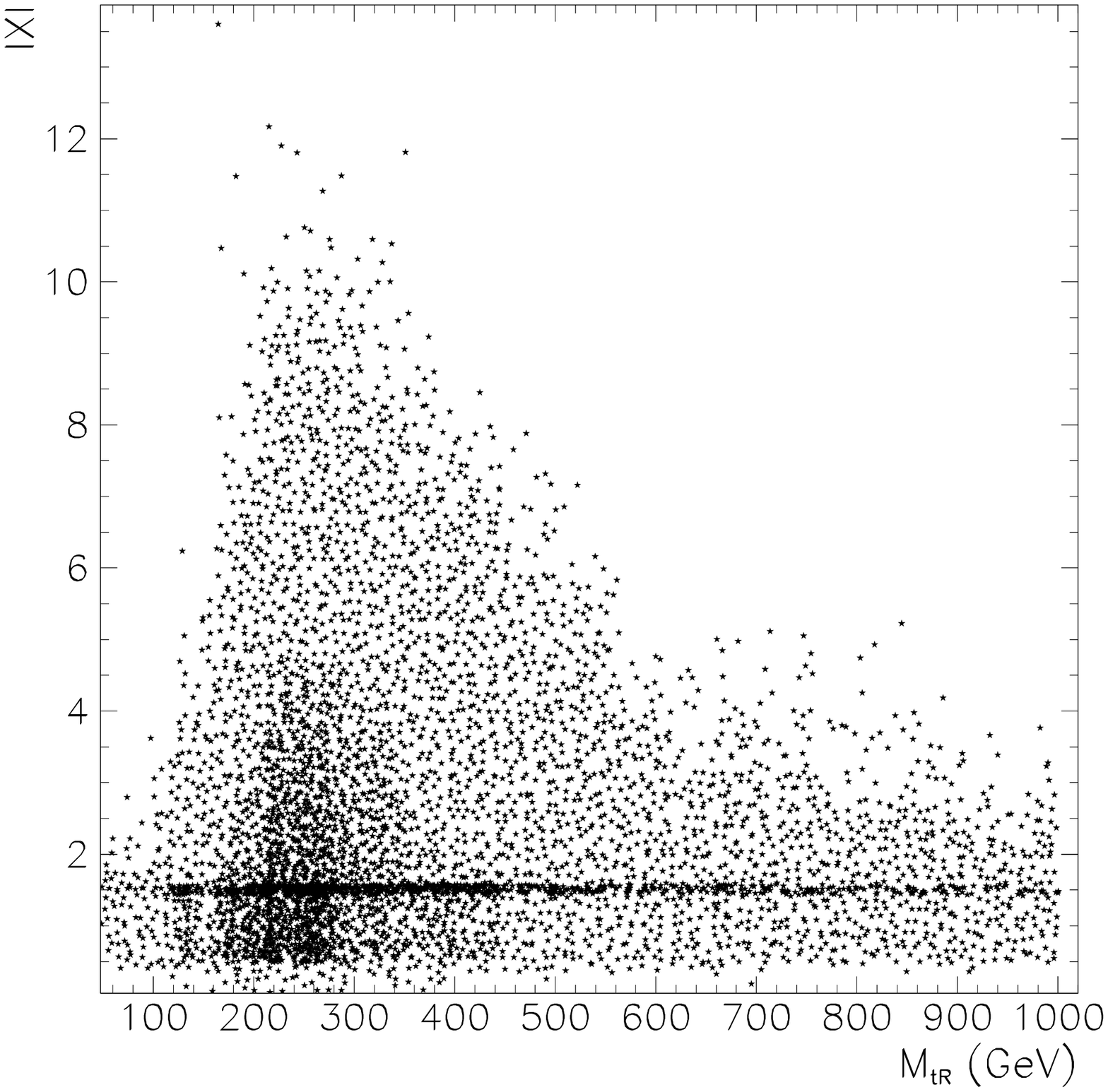,width=0.49\linewidth}
\\
\\
\epsfig{file=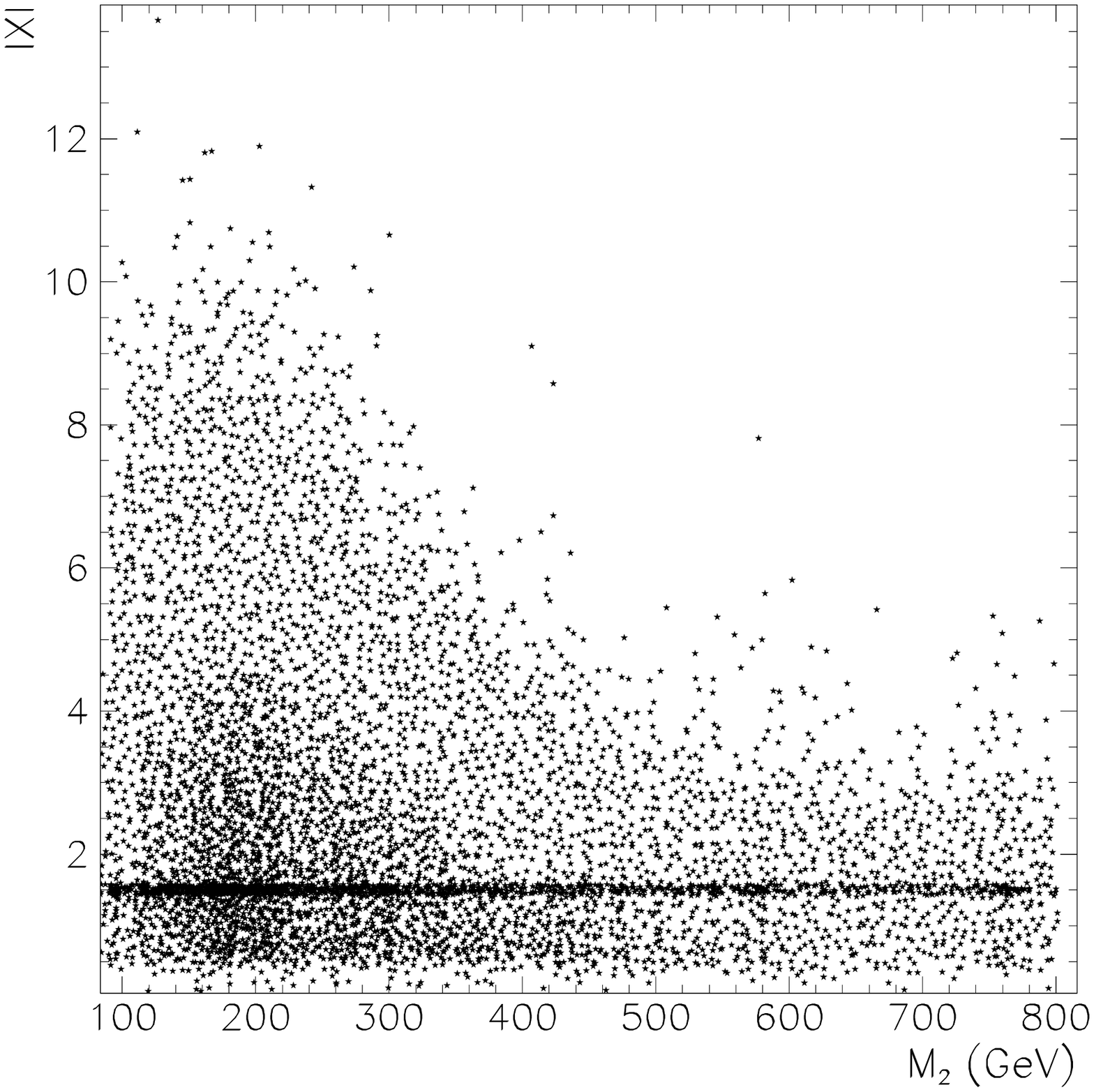,width=0.49\linewidth}
&
\epsfig{file=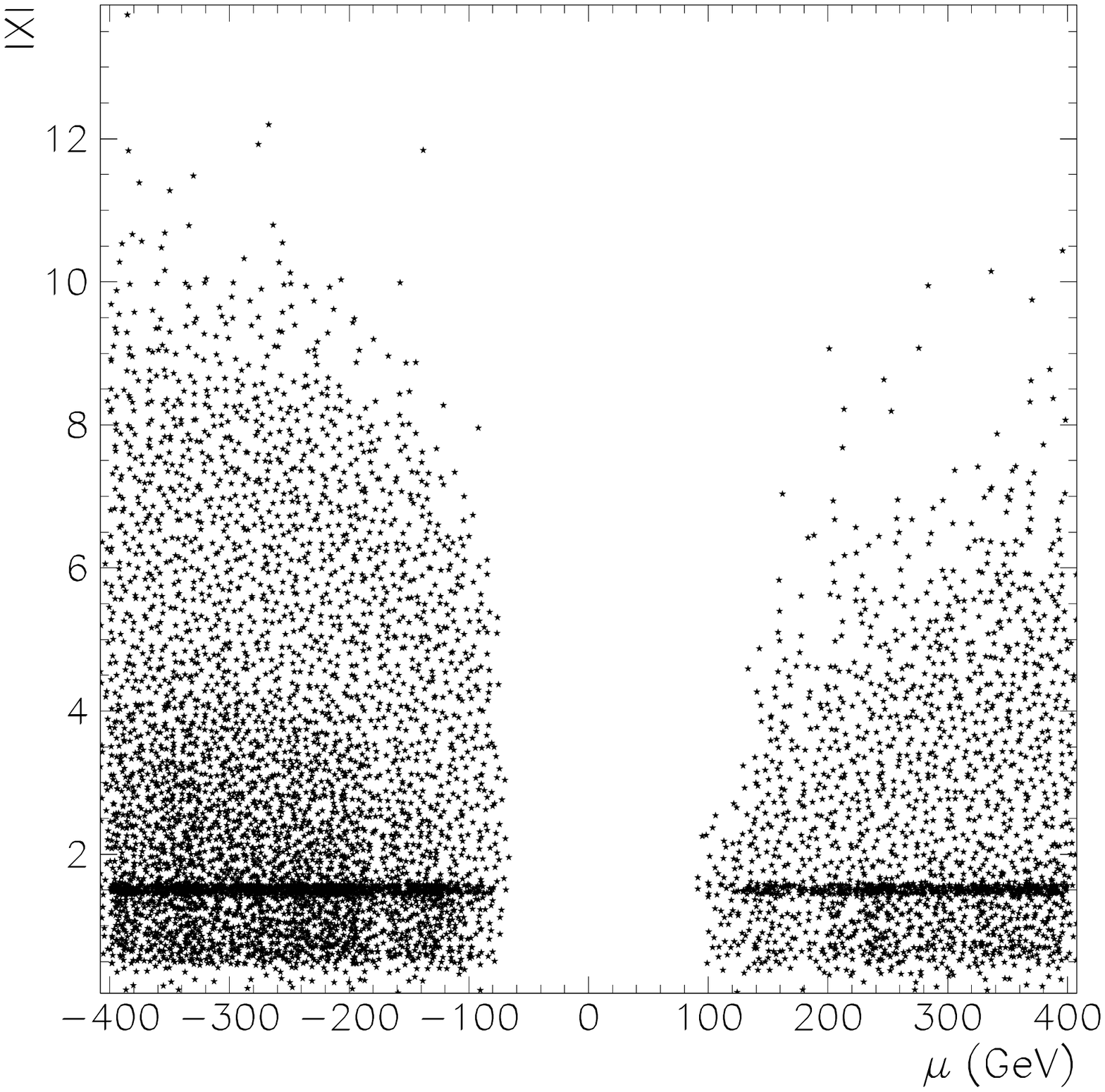,width=0.49\linewidth}
\\
\end{tabular}
\caption{Distributions of $X$ versus $\gamma$, 
$M_{\tilde{t}_R}$, $M_2$ and $\mu$ for $\tan\beta=2$ in data
set generated during VEGAS integration.}
\label{fig:vegdist}
\end{center}
\end{figure}

As a first step of our analysis, we plotted distributions of $|X|$
obtained in the ``constrained'' scan (described in detail in
Section~\ref{subsec:par}) versus various flavour diagonal and
off-diagonal parameters, in order to check the sensitivity of the
considered decay rates to them.  The plots obtained support our conjecture
that only a modest subset of the MSSM parameters is in the
first approximation relevant for our analysis, even after including
the complicated set of additional bounds.  Most of the obtained
distributions for flavour diagonal parameters are flat (dependence on
$M_A$, $M_1$, $m_{\tilde{g}}$) or almost flat (dependence on $M_{sl}$, $A$).
A more pronounced dependence could be observed only in the case of
$\gamma$ and the SUSY parameters plotted in Fig.~\ref{fig:vegdist}.

\begin{figure}[htbp]
\begin{center}
\begin{tabular}{cc}
\epsfig{file=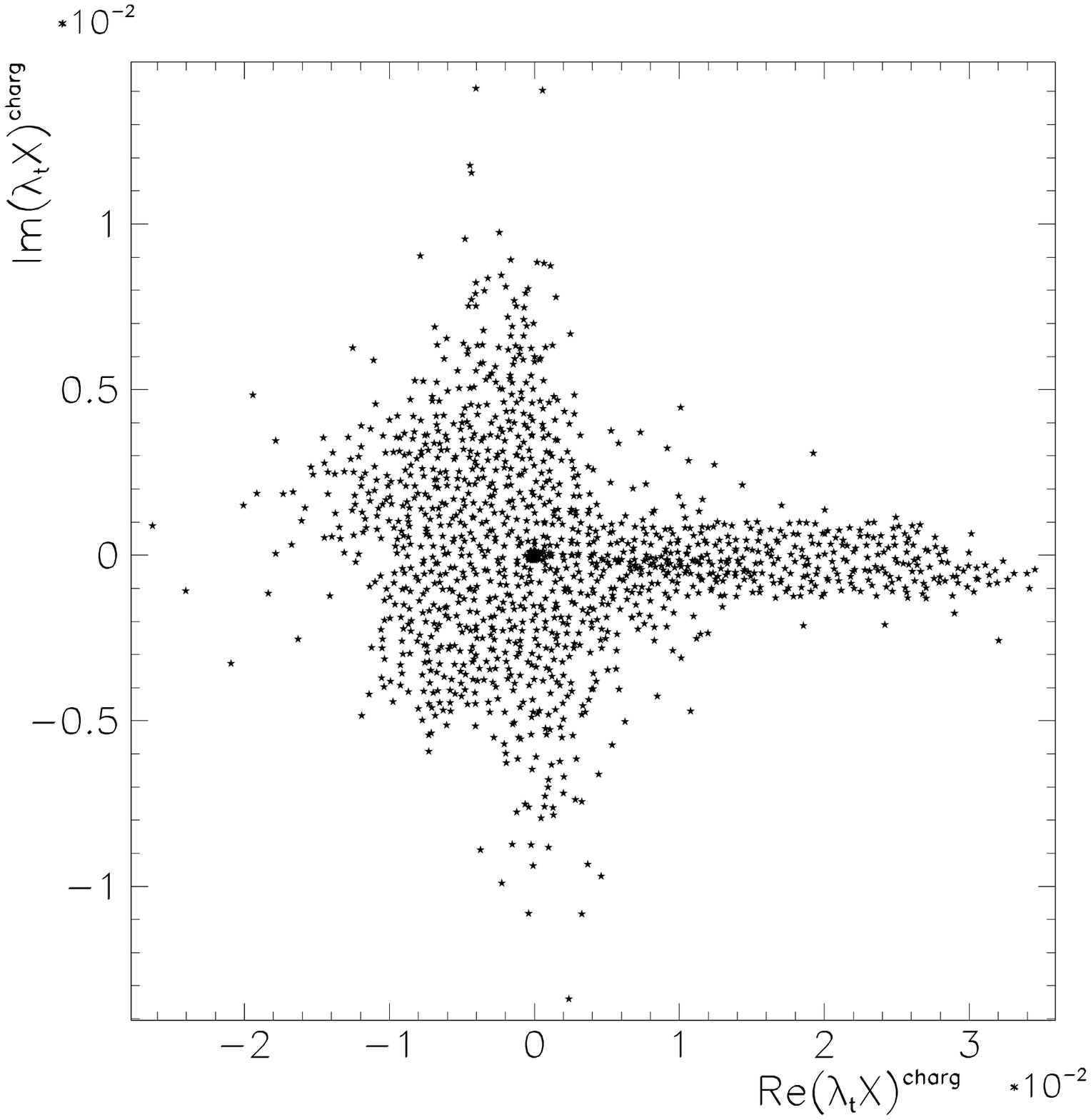,width=0.48\linewidth}
&
\epsfig{file=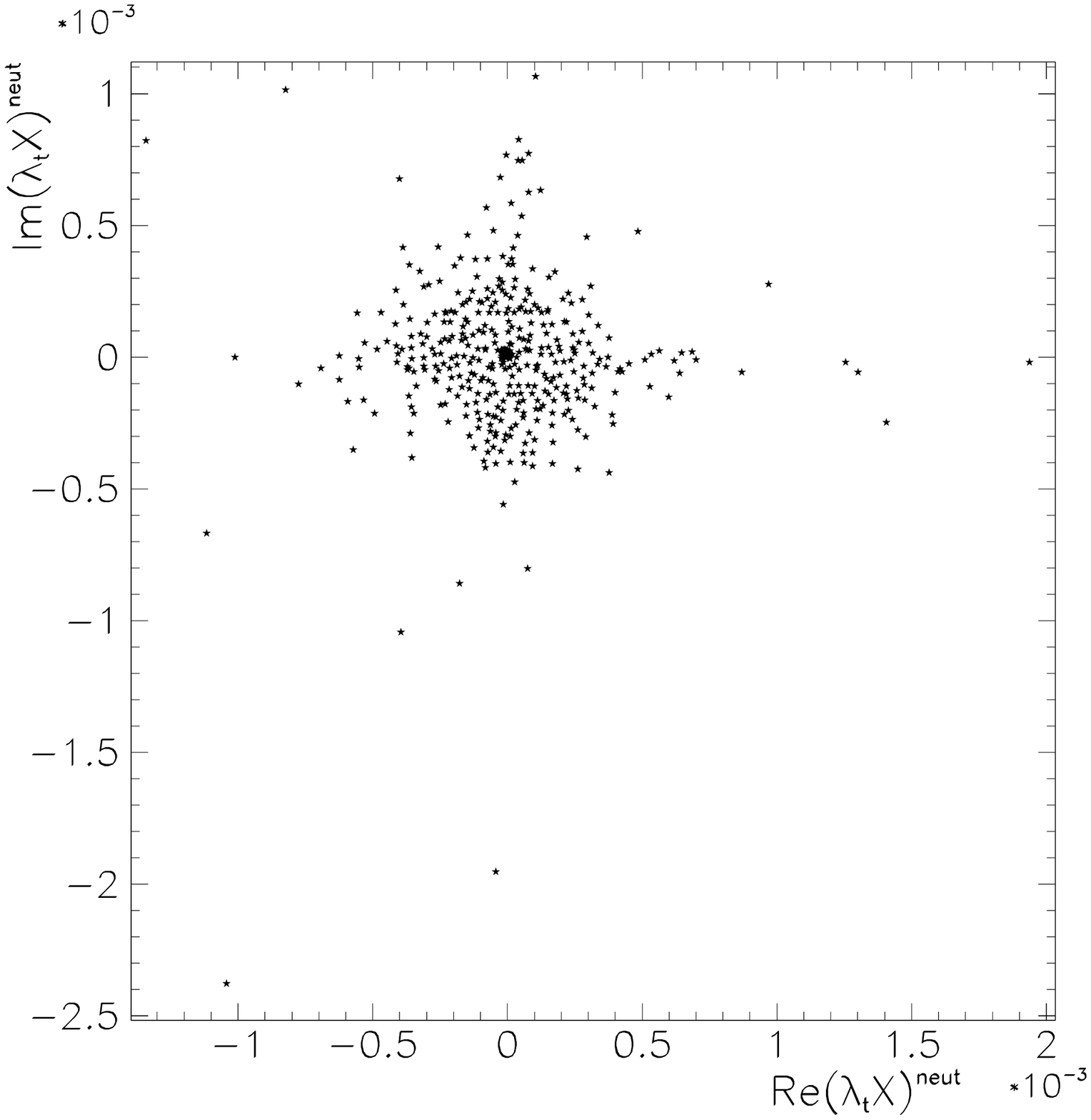,width=0.48\linewidth}
\\
\end{tabular}
\caption{Chargino and neutralino  contributions to $\lambda_t X$ 
for $\tan\beta=2$.}
\label{fig:hamdist}
\end{center}
\end{figure}

The dependence visible for the other parameters can be explained by
looking at Fig.~\ref{fig:hamdist}, where we plotted the
contributions from the chargino and neutralino sectors to $\lambda_t
X$. The chargino diagrams are always strongly dominant, typically one
order of magnitude larger than the neutralino contributions.  Gluino
exchanges (in our case penguin diagrams only) can be always completely
neglected, as we checked that they are several orders of magnitude
smaller than other contributions. Thus, the size of both decay rates,
$Br(K^+\ra \pi^+ \nu \bar \nu)$ and $Br(K_L\ra \pi^0 \nu \bar \nu)$,
is determined by the chargino-up-squark contribution and should depend
mostly on the parameters entering the expression for this amplitude,
in agreement with the results of~\cite{BRS,COLISI,Buras:1999da}.
  This is
not true for the variety of experimental and theoretical bounds we
took into account in constraining the MSSM parameter space -- they also depend
on other parameters (e.g., the hole in the $\mu$ distribution
comes of course from the bounds on the lightest neutralino and chargino
masses). This has some secondary influence on the shape of the plots in
Fig.~\ref{fig:vegdist}, as the imposed constraints can lead to correlation
between the allowed ranges of some of the parameters directly
relevant for the $K\ra \pi \nu \bar \nu$ decay calculation.  For
instance, attainable values of $|X|$ grow first with MSSM masses like
$M_{\tilde t_R}$ or $M_2$, because experimental constraints for squark
mass insertions are easier to satisfy for heavier SUSY particles
(similarly for distributions with $M_{\tilde t_L}$ or $M_{sq}$, not
included in Fig.~\ref{fig:vegdist}).  Later they go down again because the
suppression of SUSY loop diagrams contributing to $K\ra \pi \nu \bar
\nu $ decays for heavy virtual particles dominates above the effect of
weakening the impact of the experimental constraints.
One should also note that
the distribution of $\gamma$ in Fig.~\ref{fig:vegdist} shows an
increased density of points around SM-preferred value $\gamma\approx
70^{\circ}\pm 30^{\circ}$ -- for the remaining points, the larger or smaller
value of the KM phase must be compensated by the phases of some mass
insertions.

\begin{figure}[htbp]
\begin{center}
\epsfig{file=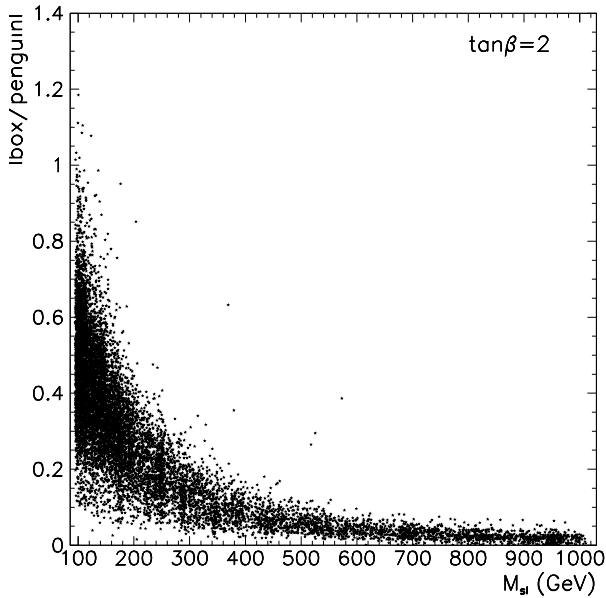,width=0.5\linewidth}
\caption{Absolute value of the ratio of box to penguin contributions as 
a function of slepton mass for $\tan\beta=2$. Only points for which
$Br(\kpn)\geq 1.5\cdot 10^{-10}$ are plotted.}
\label{fig:bprat}
\end{center}
\end{figure}

An interesting observation can be made by comparing the relative size of
penguin- and box-type contributions to the $K \ra \pi \nu \bar \nu$
decay amplitude. A common
assumption used in the literature says that box diagrams are
parameterically suppressed by ${\cal O}(M_W^2/M_{slepton}^2)$ and
can be safely neglected when compared with $Z^0$-penguin contribution.  This
is certainly not justified for light slepton masses, and not very
accurate even for moderate $M_{sl}\sim 300$ GeV. In
Fig.~\ref{fig:bprat} we plot the absolute value of the ratio of box to
penguin contribution against the slepton mass, including in the
distribution only particularly interesting points for which the $\kpn$
decay rate is large, $Br(\kpn)\geq 1.5\cdot 10^{-10}$. As can be
immediately seen, for slepton masses just above 100 GeV box and
penguin contributions are in a substantial number of cases comparable in
amplitude, and even for $M_{sl}=300$ GeV their ratio can still reach
30\%.  Thus box diagrams definitely should be taken into account in
realistic calculations. Of course, their presence introduces slepton
mass dependence into the considered branching ratios, which would
otherwise be negligible.

\begin{figure}[htbp]
\begin{center}
\begin{tabular}{cc}
\epsfig{file=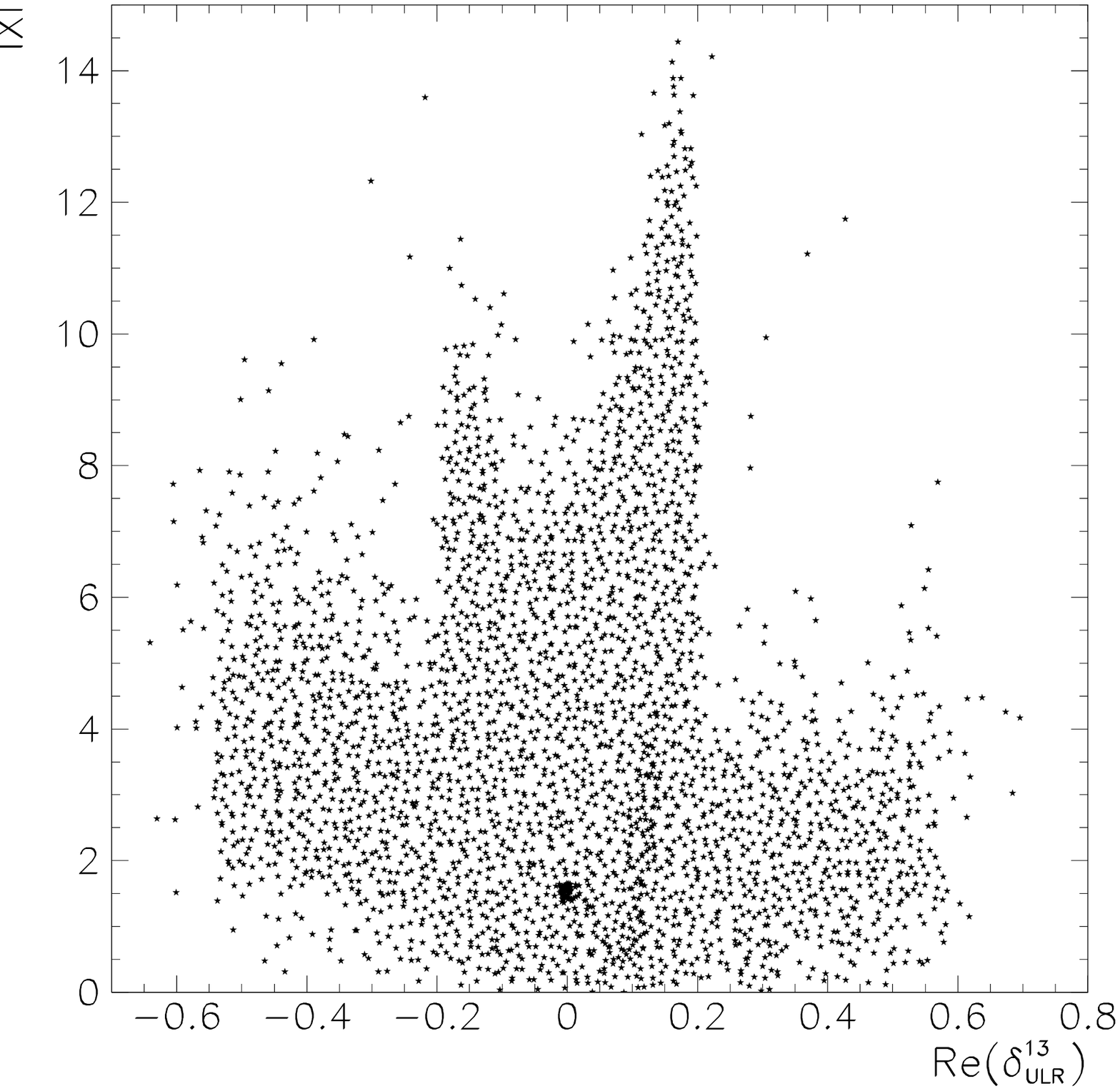,width=0.48\linewidth}
&
\epsfig{file=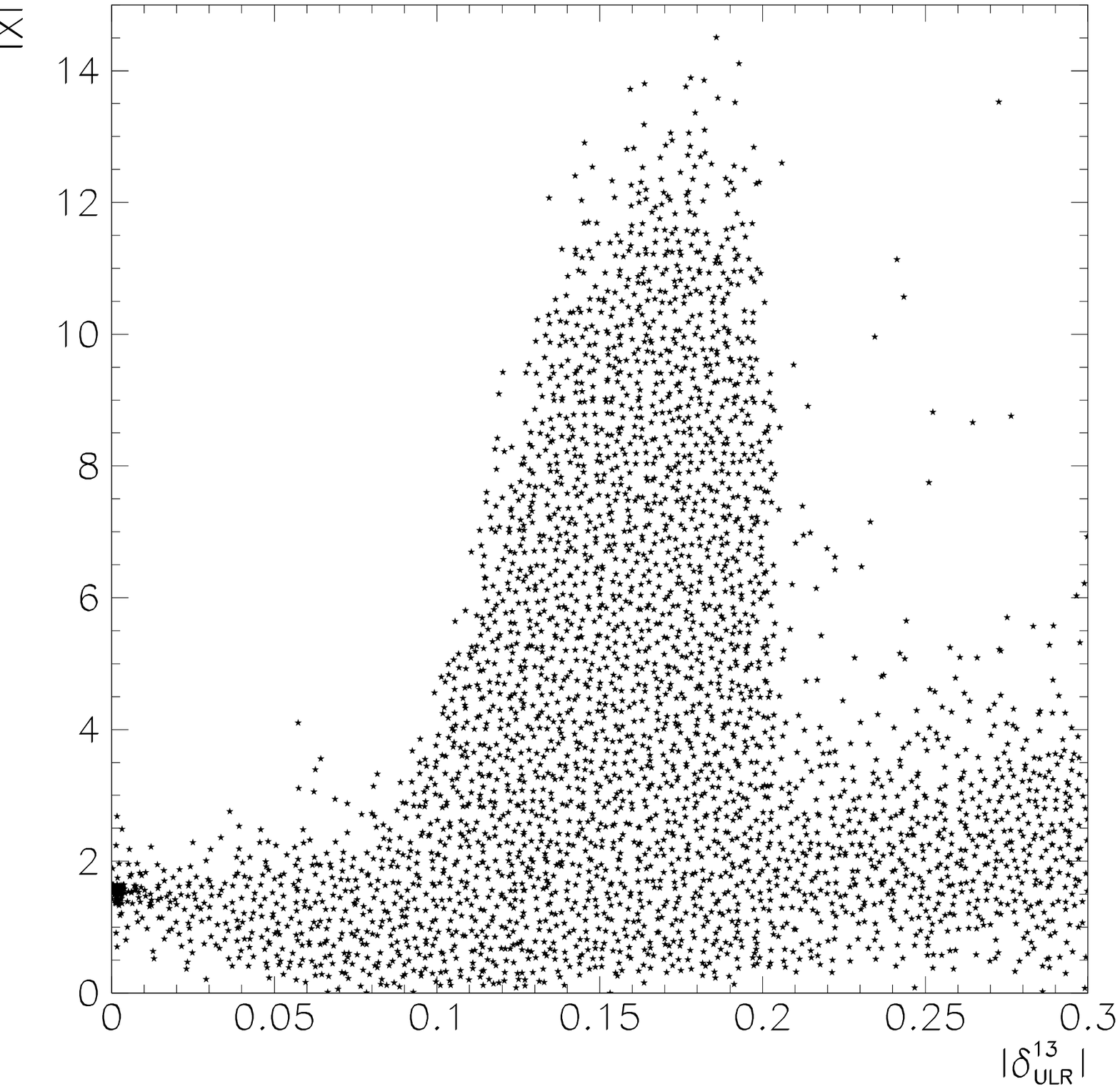,width=0.48\linewidth}
\\
\epsfig{file=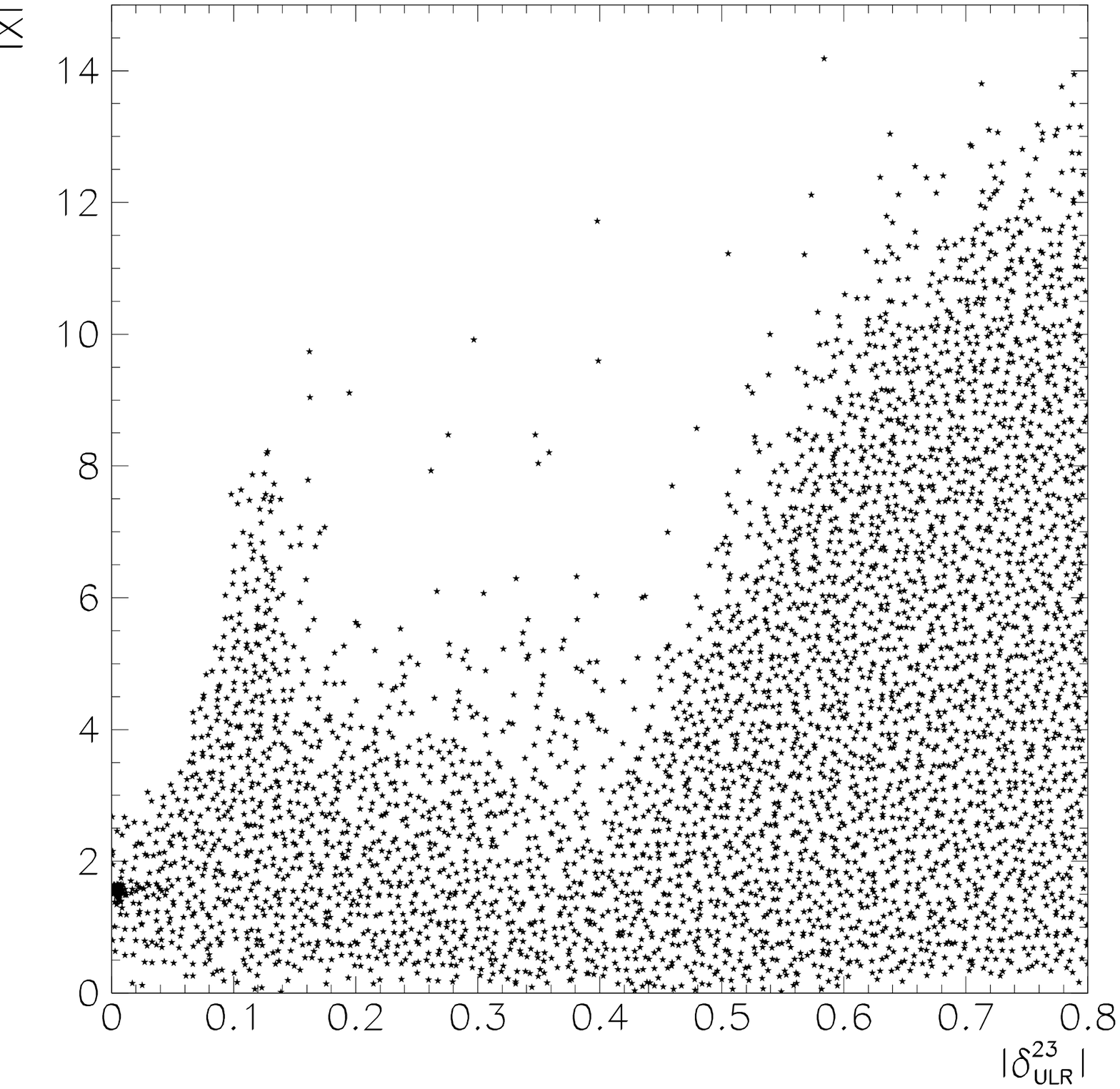,width=0.48\linewidth}
&
\epsfig{file=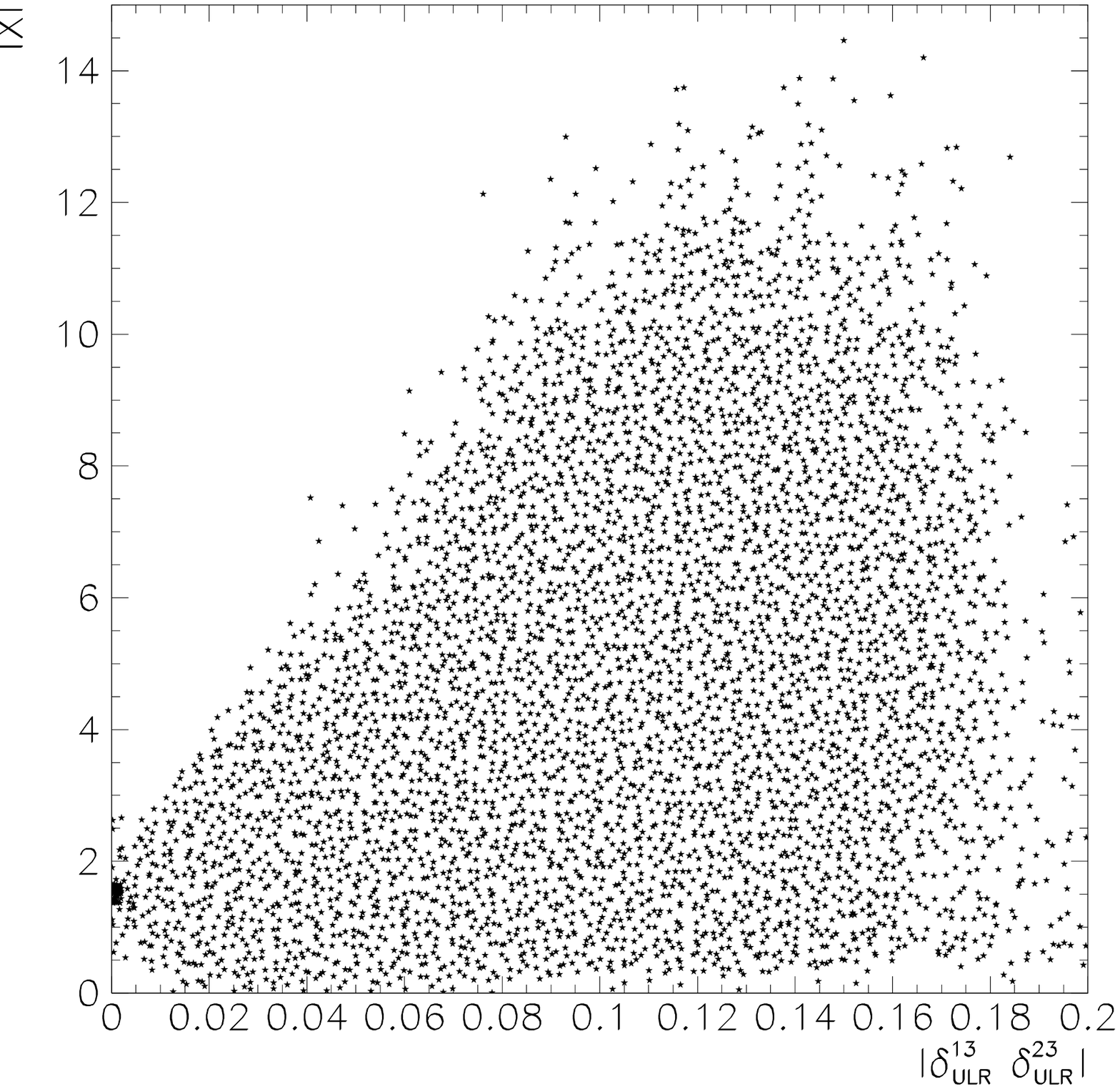,width=0.48\linewidth}
\\
\end{tabular}
\caption{Distributions of $X$ versus real part and modulus
of $\delta_{ULR}$ mass insertions for $\tan\beta=2$.
 These plots have been produced using the exact results given
in the ~\ref{app:wilson} and not the MIA.}
\label{fig:vegdist_mi}
\end{center}
\end{figure}

In our ``constrained'' scan we varied 3 types of mass insertions,
$\delta_{LL}^{12}$, $\delta_{ULR}^{13}$ and $\delta_{ULR}^{23\star}$.
We checked that the dependence on $\delta_{LL}^{12}$ is actually almost
negligible, as expected because it is strongly constrained by bounds
coming from $\epsilon_K$ and $\Delta M_K$ measurements: $\delta_{LL}^{12}\leq
0.05$. In Fig.~\ref{fig:vegdist_mi} we plot the dependence on the moduli
of the remaining two ULR mass insertions. The dependence on them is quite
pronounced, in agreement with the conclusions of~\cite{COLISI} that
the second order ULR terms in the MI expansion give the dominant
contribution to the considered decays.
We return to this point in Subsect.~\ref{subsec:litcomp} below.
For comparison, we also plot
$|X|$ against just the real part of $\delta_{ULR}^{13}$,
which exhibits less correlation.

\subsection{Allowed ranges for the \boldmath{$X$} function, 
\boldmath{$Br(K_L\ra \pi^0 \nu \bar\nu)$}  and 
\boldmath{$Br(K^+\ra \pi^+ \nu \bar \nu)$}}
\label{subsec:xdist}

\begin{figure}[htbp]
\begin{center}
\begin{tabular}{cc}
\epsfig{file=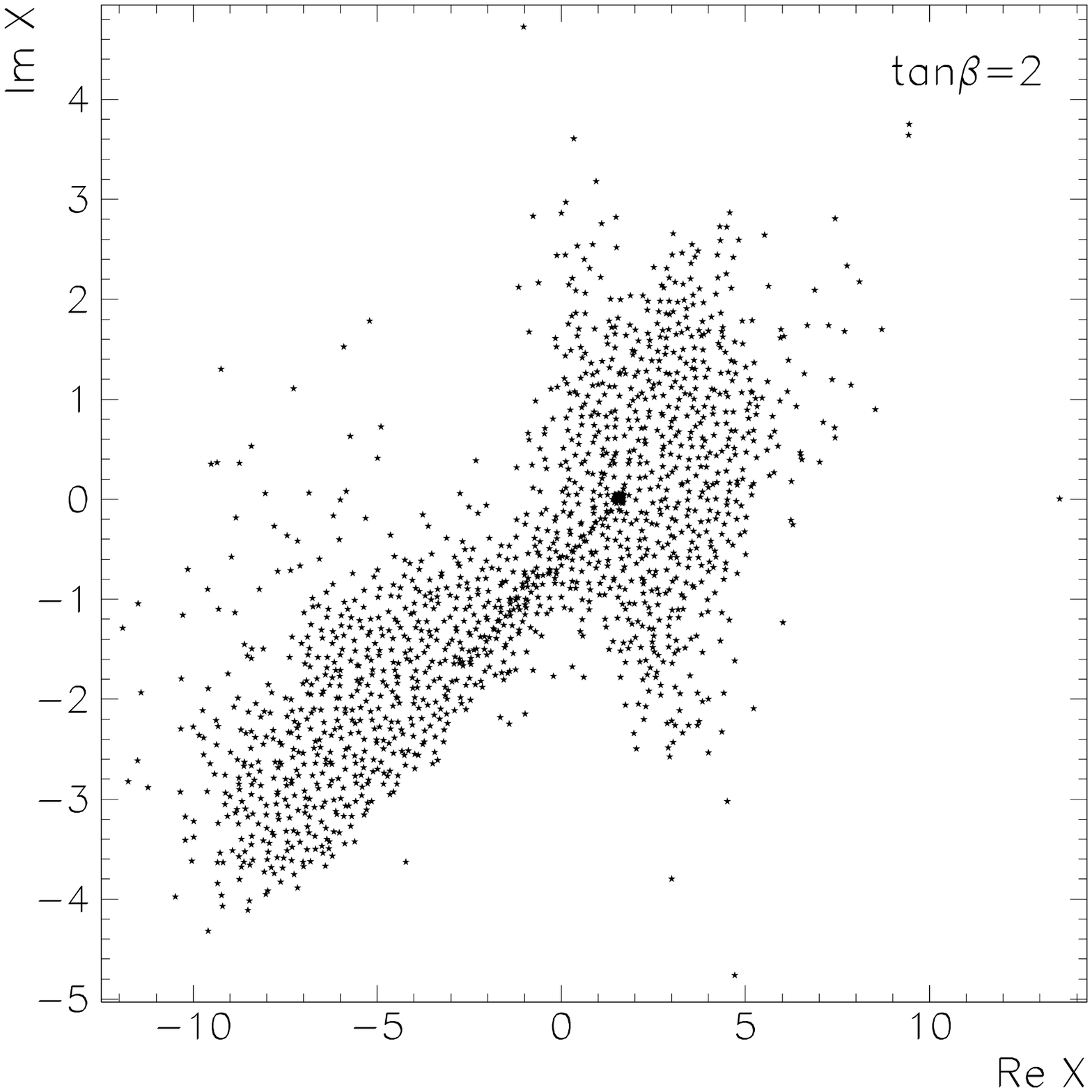,width=0.48\linewidth}
&
\epsfig{file=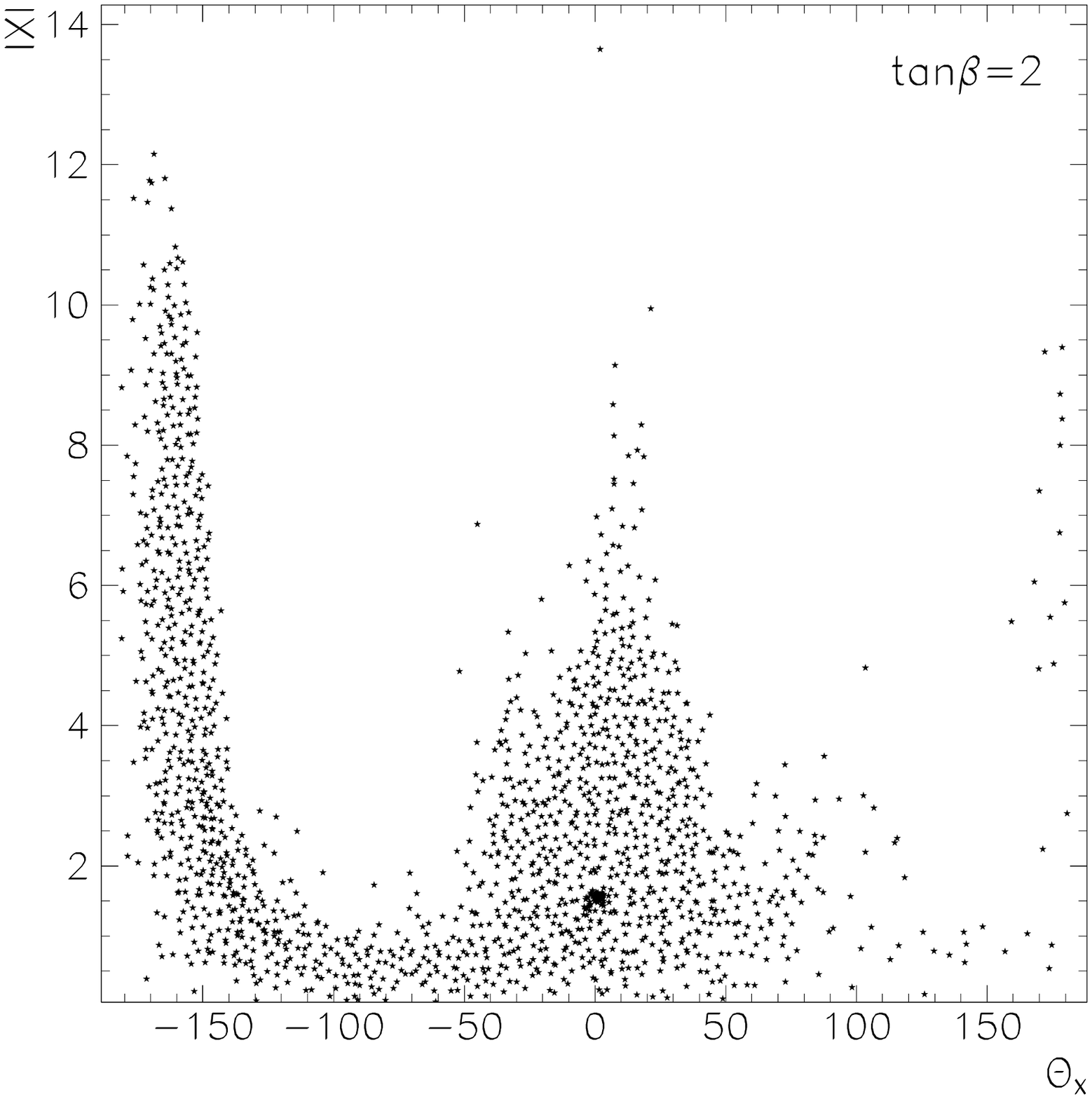,width=0.48\linewidth}
\\
\epsfig{file=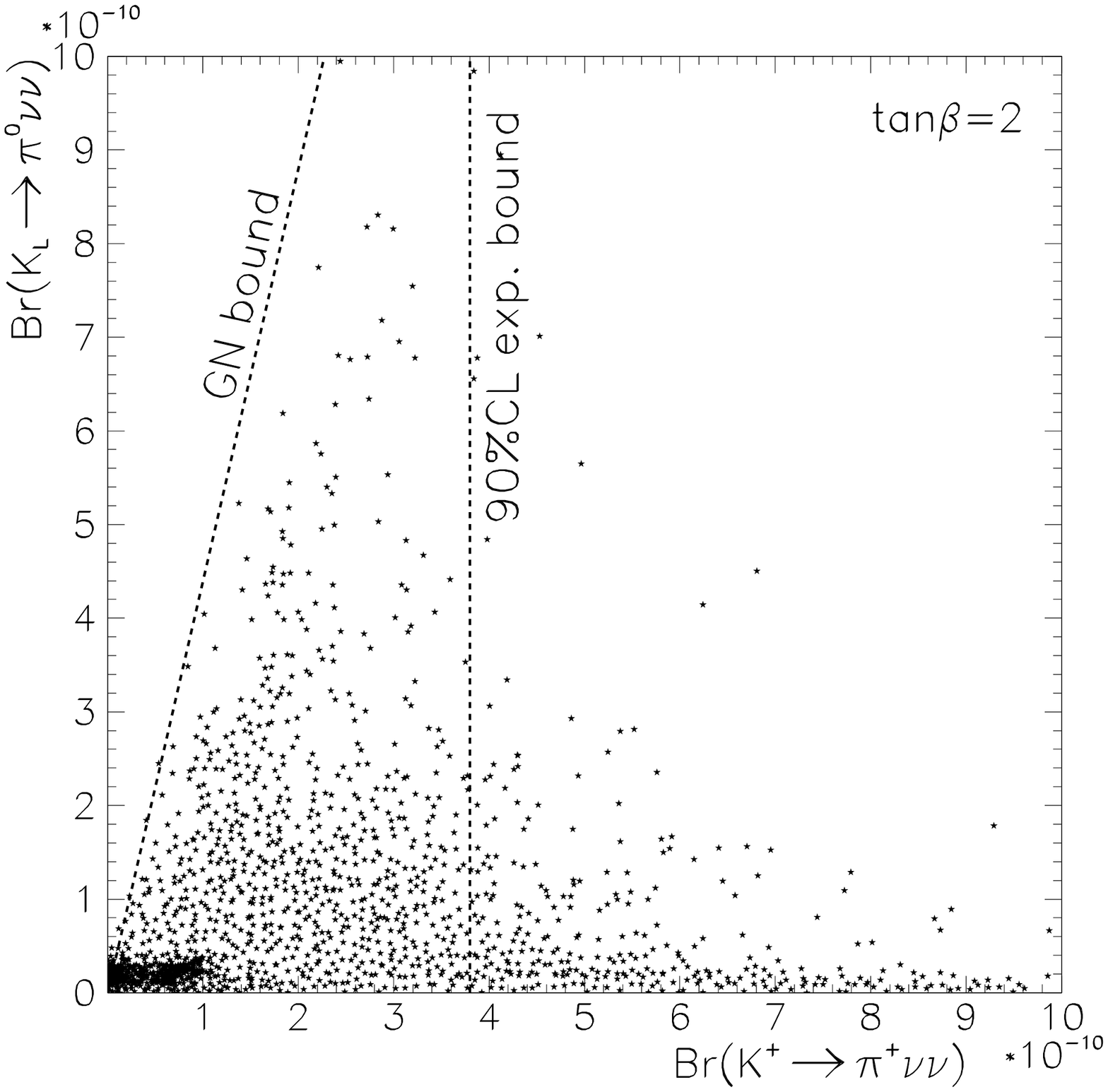,width=0.48\linewidth}
&
\epsfig{file=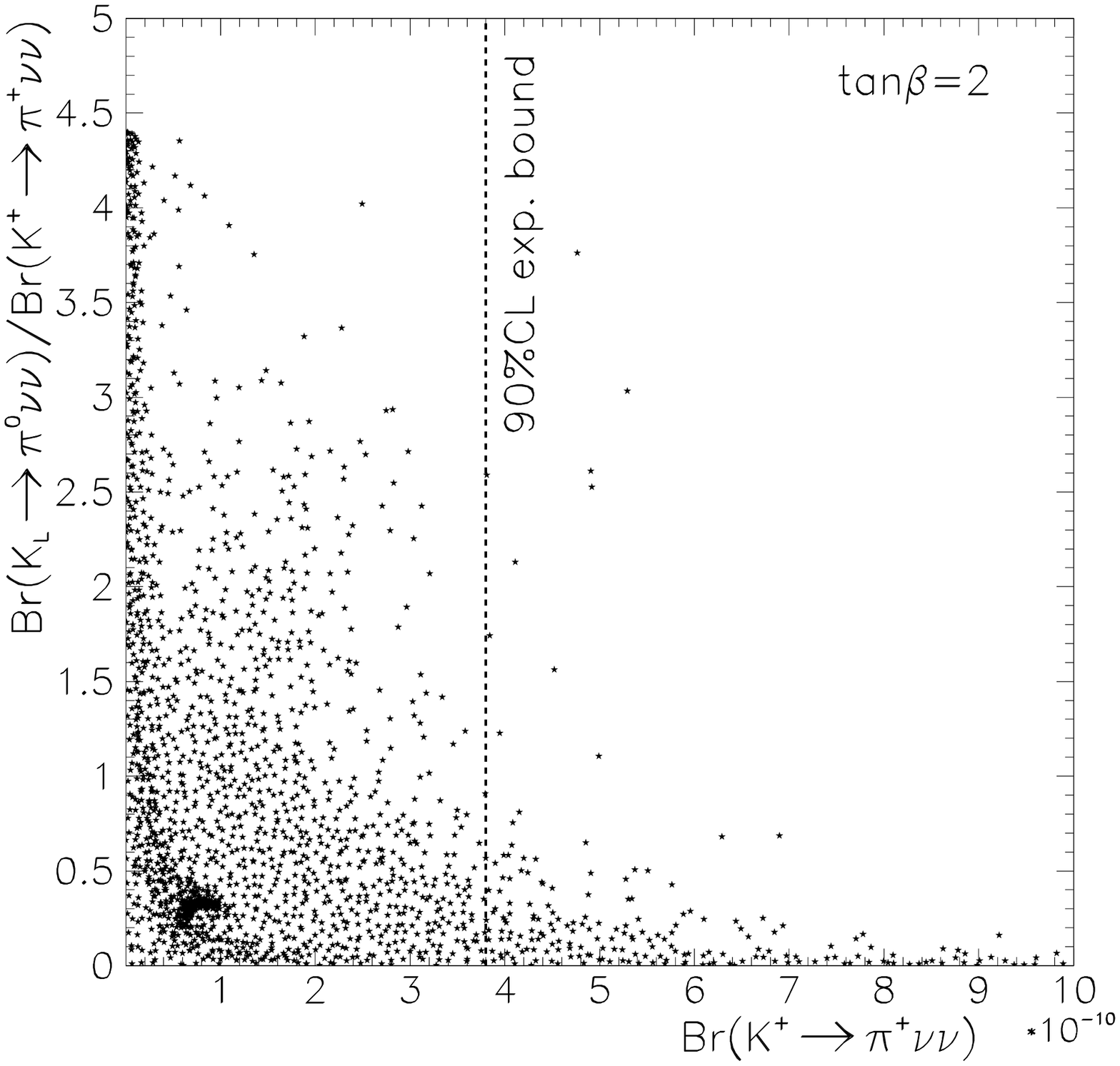,width=0.48\linewidth}
\\
\end{tabular}
\caption{Distributions of $X$ and $Br(K_L\ra \pi^0 \nu 
\bar \nu)$,  $Br(K^+\ra \pi^+ \nu \bar \nu)$ for $\tan\beta=2$.}
\label{fig:xtb2}
\end{center}
\end{figure}

The most interesting information from scanning over SUSY parameters
are the maximal values of $|X|$, $\theta_X$ and the branching
ratios $Br(K_L\ra \pi^0 \nu \bar \nu)$, $Br(K^+\ra \pi^+ \nu \bar
\nu)$ which can be obtained taking into account the bounds given by
other processes.  In Fig.~\ref{fig:xtb2} we plot distributions of
those quantities for $\tan\beta=2$.

As can be seen from Fig.~\ref{fig:xtb2}, the allowed values for $Br(K_L\ra
\pi^0 \nu \bar \nu)$ and $Br(K^+\ra \pi^+ \nu \bar \nu)$ can be 
significantly enhanced compared to the SM prediction, even by an order
of magnitude.  Such large values are higher than the 90\%CL
experimental bound~\cite{E949}, $Br(K^+\ra \pi^+ \nu \bar
\nu)<3.8\cdot 10^{-10}$, therefore this decay can already be used to
constrain the MSSM parameter space. 
We will analyze this elsewhere.
One should note that the obtained
values for the branching ratios do not violate the Grossman-Nir
(GN)~\cite{GRNIR} bound $Br(K_L\ra \pi^0 \nu \bar \nu)/Br(K^+\ra \pi^+
\nu \bar \nu) \leq 4.4$, which can be regarded as a simple cross-check of 
the correctness of our numerical codes.

\begin{figure}[htbp]
\begin{center}
\begin{tabular}{cc}
\epsfig{file=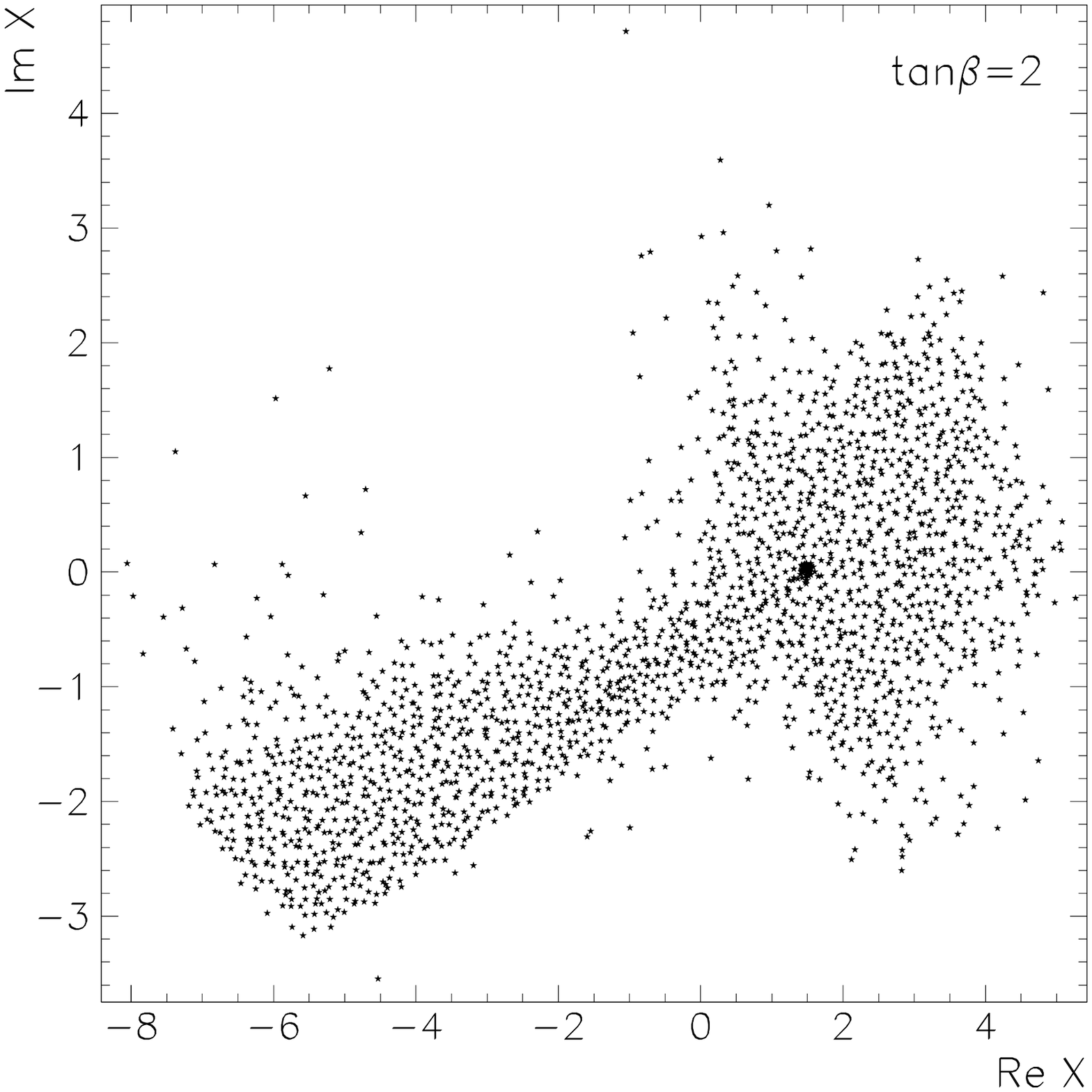,width=0.48\linewidth}
&
\epsfig{file=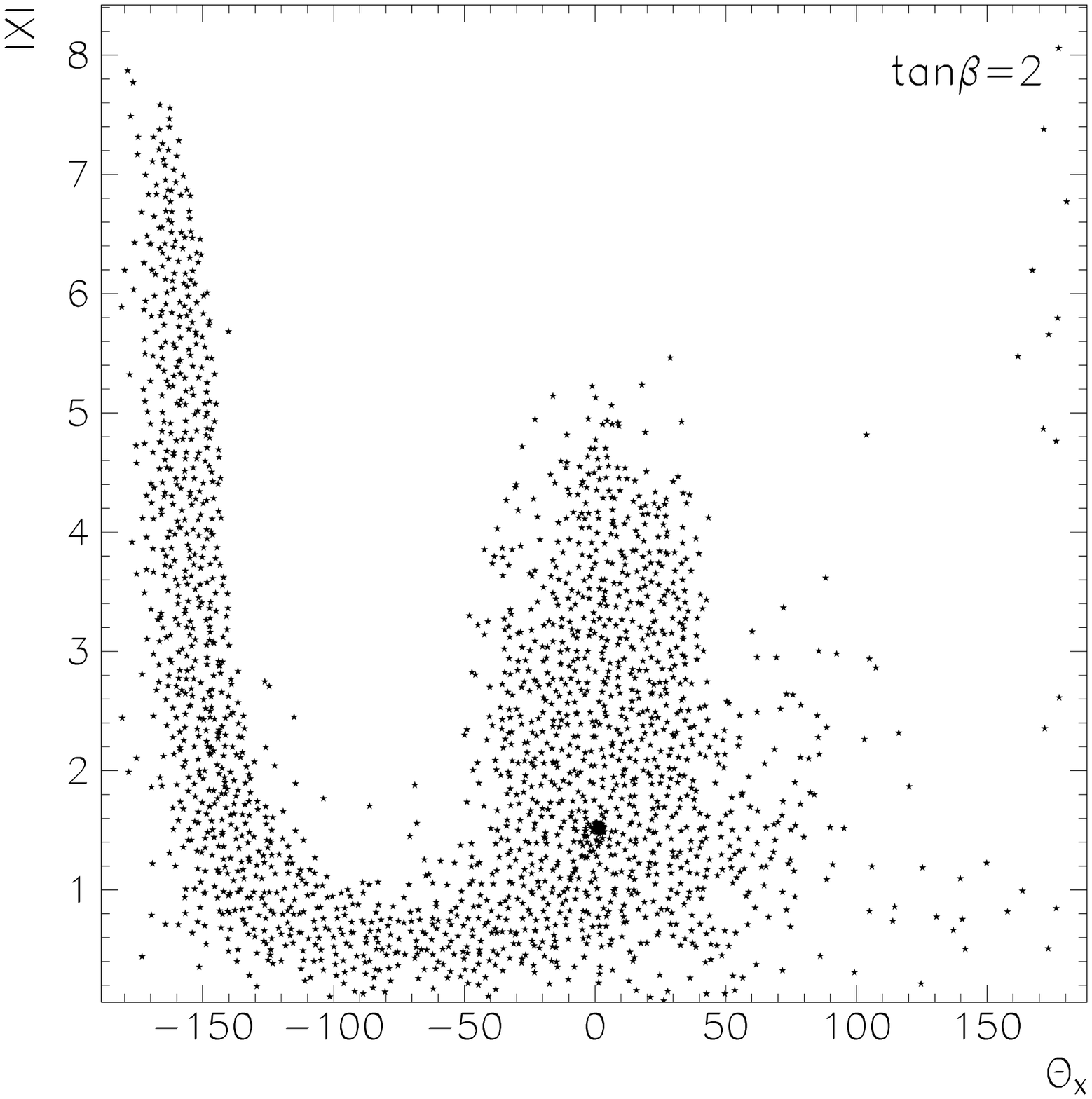,width=0.48\linewidth}
\\
\end{tabular}
\caption{Distributions of $X$ for $\tan\beta=2$ after imposing 
constraint $Br(K^+\ra \pi^+ \nu \bar \nu)<3.8\cdot 10^{-10}$.}
\label{fig:xtb}
\end{center}
\end{figure}

In Fig.~\ref{fig:xtb} we plot the allowed range of $X$ after imposing the
cut~\cite{E949} $Br(K^+\ra \pi^+ \nu \bar \nu)<3.8\cdot 10^{-10}$.
Even with this constraint, $|X|$ could be several
times bigger than the SM value~(\ref{eq:xsm}).  Also, its phase can
still vary almost freely, however is preferred to be in a broad range
$-160^{\circ}\leq \theta_X \leq 70^{\circ}$. Such a freedom leads to
a possible enhancement of the ratio $Br(K_L\ra \pi^0 \nu \bar
\nu)/Br(K^+\ra \pi^+ \nu \bar \nu)$ even for large values of
$Br(K^+\ra \pi^+ \nu \bar \nu)$, as can be observed in
Fig.~\ref{fig:xtb2}.

 The plots of $X$ in Figs.~\ref{fig:xtb2} and~\ref{fig:xtb} display
a conspicuous correlation between the phase and modulus. This can be
understood from the allowed region in the complex plane for the
chargino contribution to $\lambda_t X$ shown in Fig.~\ref{fig:hamdist}.
The experimental constraints we apply constrain its imaginary part to
be relatively small, especially when the real part becomes large. Taken
together with the fact that ${\rm Re} \lambda_t<0$ (any $\gamma$) and
that $\gamma$ is sufficiently constrained (cf. Fig.~\ref{fig:vegdist}) to
 always imply ${\rm Im} \lambda_t>0$, this
explains the shape of the allowed region in the complex $X$ plane.

\begin{figure}[htbp]
\begin{center}
\begin{tabular}{cc}
\epsfig{file=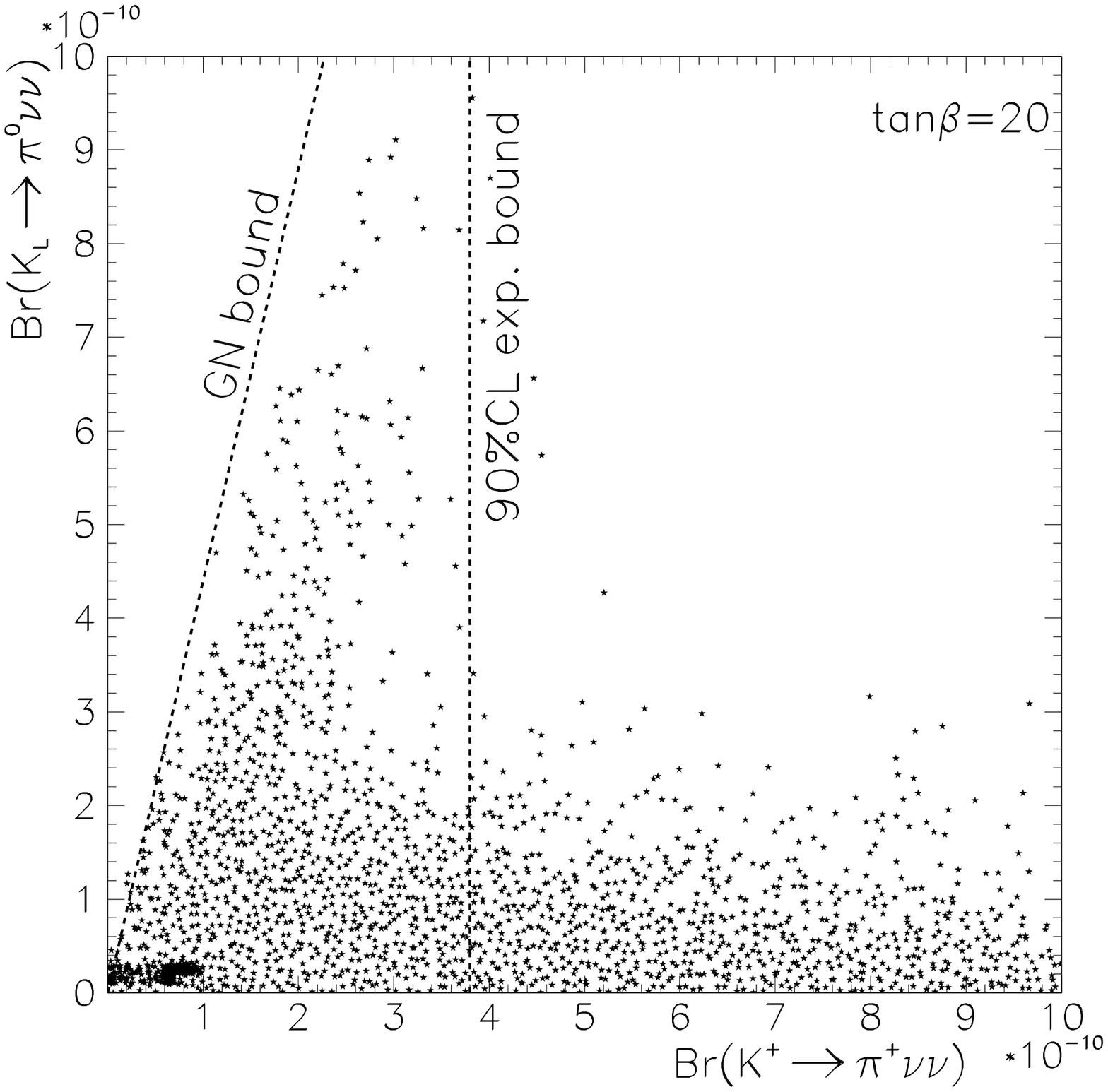,width=0.48\linewidth}
&
\epsfig{file=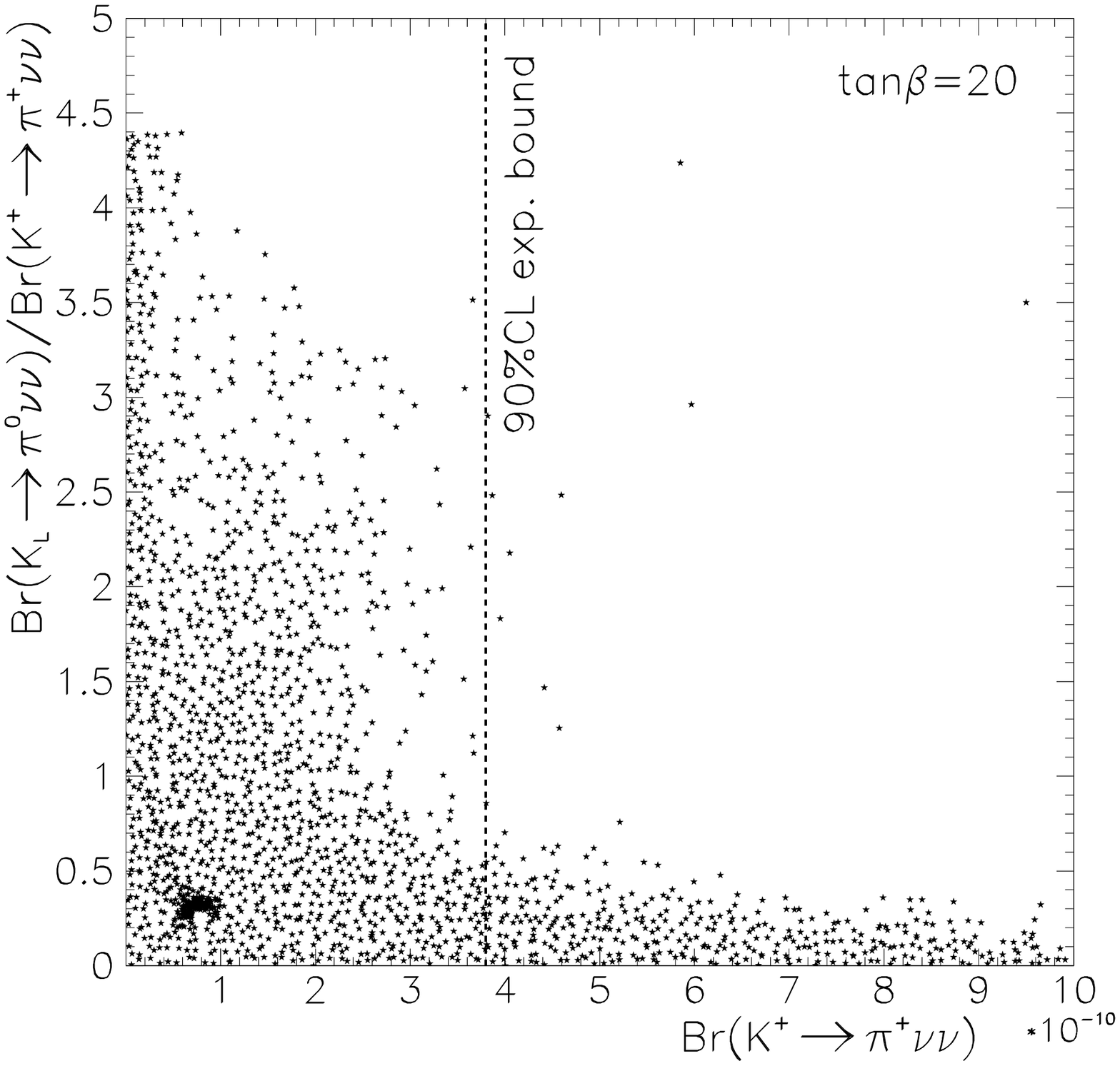,width=0.48\linewidth}
\\
\end{tabular}
\caption{Distributions of $Br(K_L\ra \pi^0 \nu \bar \nu)$ 
and $Br(K^+\ra \pi^+ \nu \bar \nu)$ for $\tan\beta=20$.}
\label{fig:xtbp}
\end{center}
\end{figure}

Finally, in Fig.~\ref{fig:xtbp} we plot distributions of $Br(K_L\ra
\pi^0 \nu \bar \nu)$ and $Br(K^+\ra \pi^+ \nu \bar \nu)$  for higher 
value of $\tan\beta=20$.  Results are qualitatively similar to those
obtained for $\tan\beta=2$, however for larger $\tan\beta$
(particularly $\tan\beta\geq 10$) it is easier to generate parameters
sets giving high branching ratios (or their ratio).

To better illustrate the effects shown in the figures in this
section, we give in Table~\ref{tab:example} specific examples
of MSSM parameter choices leading to particularly interesting results.
The quotation marks on some quantities in this table indicate that in
evaluating these quantities we have set hadronic parameters to their central 
values as discussed in Subsection~\ref{subsec:bounds}. Varying these
parameters, it is easy to reproduce the relevant experimental 
data within one standard deviation. 

\begin{table}[htbp]
\begin{center}
\begin{tabular}{ccc}
Parameter & Example 1 & Example 2 \\ \hline
$\tan\beta$ & 2 & 20 \\
$M_A$ & 333 & 260 \\
$\mu$ & -375 & -344 \\
$M_{\tilde g}$ & 437 & 928 \\ 
$M_2$ & 181 & 750 \\
$M_{sq}$ & 308 & 608 \\
$M_{\tilde t_L}$ & 138 & 215 \\
$M_{\tilde t_R}$ & 279 & 338 \\
$M_{sl}$ & 105 & 884 \\
$A$ & -0.289 & -0.342 \\
$\gamma$ & $64^{\circ}$ &  $38^{\circ}$ \\
$\delta_{LL}^{12}$ & $(2.18-5.02i)\cdot 10^{-5}$& $(7.57-0.871i)\cdot 10^{-4}$\\
$\delta_{ULR}^{13}$ & $(-1.52+0.748i)\cdot 10^{-4}$ & $0.292-0.213i$ \\
$\delta_{ULR}^{23\star}$ & $0.001-0.604i$ & $0.239-0.195i$ \\
``$|\varepsilon_K|$" & $2.35\cdot 10^{-3}$ & $2.10\cdot 10^{-3}$ \\
``$\Delta M_d$" & $3.15\cdot 10^{-13}$ & $2.55\cdot 10^{-13}$ \\
``$\Delta M_s$" & $1.03\cdot 10^{-11}$ & $1.19\cdot 10^{-11}$ \\
``$Br(B\ra X_s\gamma)$" & $3.88\cdot 10^{-4}$ & $3.93\cdot 10^{-4}$ \\
$Br(\kpn)$ & $1.78\cdot 10^{-10}$ & $2.07\cdot 10^{-10}$ \\
$Br(\klpn)$ & $3.08\cdot 10^{-11}$ & $4.34\cdot 10^{-10}$ \\
\end{tabular}

\caption{\label{tab:example}
Examples of MSSM parameter points passing experimental constraints at
the assumed accuracy (see discussion in section~\ref{subsec:bounds})
and giving enhanced $\kpn$ and $\klpn$ decay rates.
Units are as in Table~\ref{tab:constr}. See text for the meaning 
of quotation marks. }
\end{center}
\end{table}

\subsection{Comparison with earlier literature}
\label{subsec:litcomp}

Colangelo and Isidori~\cite{COLISI}, extending and clarifying
earlier work~\cite{NIRWOR,BRS}, have used the fact that all contributions
to the $Z^0$-penguin must be at least of second order in the VEV by
gauge invariance to trace the dominant SUSY contribution, at least in
the case of large $X$, to the following term, which
can be obtained as part of the chargino contribution~(\ref{eq:PLc})
to the $Z^0$-penguin:
\be             \label{eq:PLc0}
        \left[P_L^{c(0)}\right]_{JI} = - \frac{1}{8} V_{RI}^{} V_{SJ}^*
                   Z_U^{Rk*} Z_U^{(N+3)k} Z_U^{(N+3)i*} Z_U^{Si}
                     Z_+^{1m} Z_+^{1m*} C_2(m^2_{C_m},m^2_{U_k},m^2_{U_i}) .
\ee
In the mass insertion approximation, this term contributes only
starting from the second order,
where it is proportional to the product $\delta_{ULR}^{13} \delta_{ULR}^{23*}$.

Fig.~\ref{fig:vegdist_mi} is consistent with this 
observation \cite{COLISI},
as is evidenced by the triangular excluded region in the bottom-right plot
in that figure:  $|X|$ is only large when also 
$|\delta_{ULR}^{13} \delta_{ULR}^{23*}|$ is large.

In fact, together with the large box contributions
we reported in Subsect.~\ref{subsec:sens} and Fig.~\ref{fig:bprat},
this plot suggests that even the box contributions, which were found to
be subleading in~\cite{COLISI} and earlier in~\cite{NIRWOR}, follow the
same pattern, a point that has so far not been made in the literature.
Indeed, there is no reason why the box diagram should
not be competitive with the up-squark-chargino contribution to the
$Z^0$-penguin for light slepton mass.
The chargino-box contribution~(\ref{eq:BLc}),
ignoring slepton flavour violation and charged-lepton masses, contains a
term
\begin{eqnarray}             \label{eq:BLc0}
        \left[B_L^{c(0)}\right]_{JIKL} &=&
                - \displaystyle\frac{M_W^2}{2} \delta_{KL} V_{RI}^{} V_{SJ}^*
                   Z_U^{Rk*} Z_U^{Sk} Z_-^{1m} Z_-^{1n*} Z_+^{1m} Z_+^{1n*}\nonumber\\[1mm]
                &\times& m_{C_m} m_{C_n} D_0(m^2_{C_m},m^2_{C_n},m^2_{U_k},m^2_{L_L}) ,
\end{eqnarray}
which is quite similar to~(\ref{eq:PLc0}).
Consequently, it receives a similar second-order
contribution in the mass insertion approximation,
in contrast to the finding in~\cite{COLISI} that the box diagrams
contribute only at the level of three 
off-diagonal mass insertions.
The double squark left-right flip is again the most important mixing parameter,
even though no $SU(2)$-breaking powers
of the VEV are required, unlike the case of the $Z^0$-penguin.
For the chargino contribution, that requirement therefore turns out to be a red
herring, although it still explains why gluino and neutralino-penguin
contributions are small.
All large effects are proportional to the double up-squark left-right
mixings simply because these are weakly constrained and at the same
time avoid the double CKM suppression, both of which facts were correctly
stated by the authors  of \cite{COLISI}.

More stringent bounds on the left-right mass insertions than those used in
this work have been derived in~\cite{Buras:1999da} from a
general MSSM RG analysis,
however their results depend on the assumption that the soft-breaking terms
are present already at the GUT scale. This is a model-dependent assumption,
not necessarily true in the general MSSM. There are also potentially dangerous
contributions from double mass insertions to $\Delta S=2$ processes that
might lead to a violation of bounds from $K^0 - \bar K^0$ mixing.
These are automatically taken into account in our computation
by our exact diagonalization of the squark mass matrices.

\subsection{Extended scan over MSSM parameters}
\label{subsec:full}

The ``adaptive scan'' method~\cite{BREIN} desribed in
Section~\ref{subsec:scanadapt} allows for efficient exploration of
really huge multi-dimensional parameter spaces, especially when the
dependence of the analyzed results on most of those parameters is not
very strong.  Therefore, we tried to check how our results change if
we get rid of virtually all assumptions usually used to relate MSSM
parameters and treat them all as free independent quantities.  We
varied randomly the following quanties:
\begin{itemize}
\item the angle $\gamma$  (real)
\item CP-odd Higgs mass $M_A$ (real)
\item $U(1)$ gaugino mass $M_1$ (complex)
\item $SU(2)$ gaugino mass $M_2$ (complex)
\item gluino mass $m_{\tilde{g}}$ (real)
\item $\mu$ parameter (complex)
\item diagonal left slepton mass $m^2_L$, common for all generations 
(real)
\item diagonal right slepton mass $m^2_R$, common for all generations 
(real)
\item 9 independent diagonal mass parameters in squark mass matrices, 
3 parameters for each of left, up-right and down-right mass matrix
(all real)
\item common sfermion LR mixing parameter $A$ (real) 
\item 3 independent LL mass insertions in squark mass matrices: 
$\delta_{LL}^{12}$, $\delta_{LL}^{13}$, $\delta_{LL}^{32}$ (all
complex)
\item 6 independent RR mass insertions in squark mass matrices: 
$\delta_{DRR}^{12}$, $\delta_{DRR}^{13}$, $\delta_{DRR}^{32}$,
$\delta_{URR}^{12}$, $\delta_{URR}^{13}$, $\delta_{URR}^{32}$ (all
complex)
\item 12 independent LR up- and down-squark mass insertions 
$\delta_{DLR}^{12}$, $\delta_{DLR}^{13}$, $\delta_{DLR}^{32}$,
$\delta_{DLR}^{21}$, $\delta_{DLR}^{31}$, $\delta_{DLR}^{23}$,
$\delta_{ULR}^{12}$, $\delta_{ULR}^{13}$, $\delta_{ULR}^{32}$,
$\delta_{ULR}^{21}$, $\delta_{ULR}^{31}$, $\delta_{ULR}^{23}$ (all
complex)
\end{itemize}
Altogether this gives 63 real degrees of freedom, more than half of the
free parameters of the completely unconstrained (but $R$-parity-conserving)
MSSM.  For the SUSY mass parameters we use the same limits
as given in Table~\ref{tab:constr}, with the additional
requirement $|M_1|\geq 20$ GeV;
for new complex parameters we assume their phase to vary completely
freely.  Of course, for any parameter point we still apply all
experimental and theoretical constraints listed in
Section~\ref{subsec:bounds}.  Such an extensive scan was probably
never reported before in the literature, so it is very interesting
even to see if such a general MSSM version still retains any predictive
power. The answer is positive -- the results obtained for both analyzed
decays, $K_L\ra \pi^0 \nu \bar \nu$ and $K^+\ra \pi^+
\nu \bar \nu$, remain qualitatively similar to those already discussed
for the ``constrained'' scan case!

\begin{figure}[htbp]
\begin{center}
\begin{tabular}{cc}
\epsfig{file=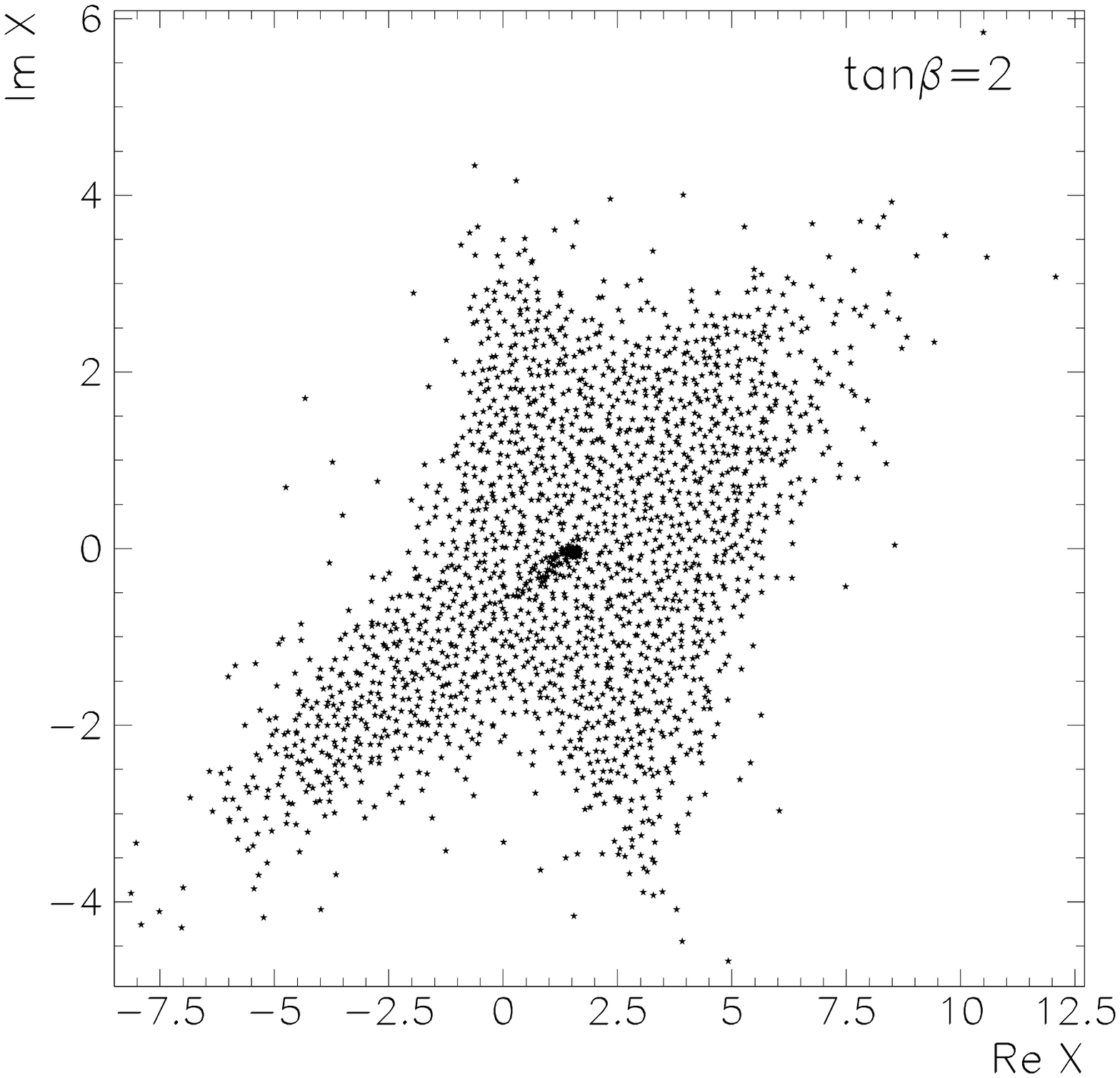,width=0.48\linewidth}
&
\epsfig{file=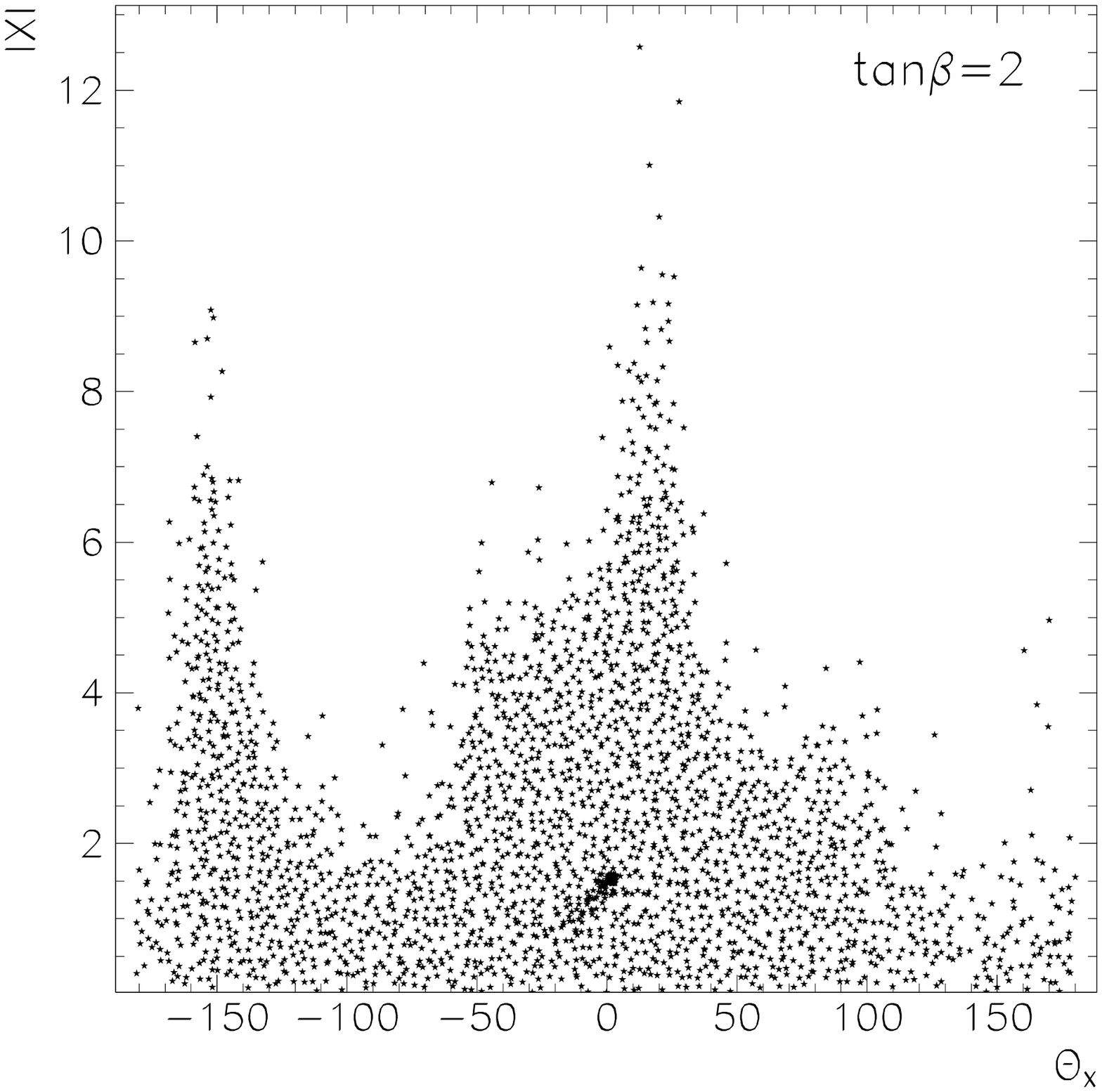,width=0.48\linewidth}
\\
\epsfig{file=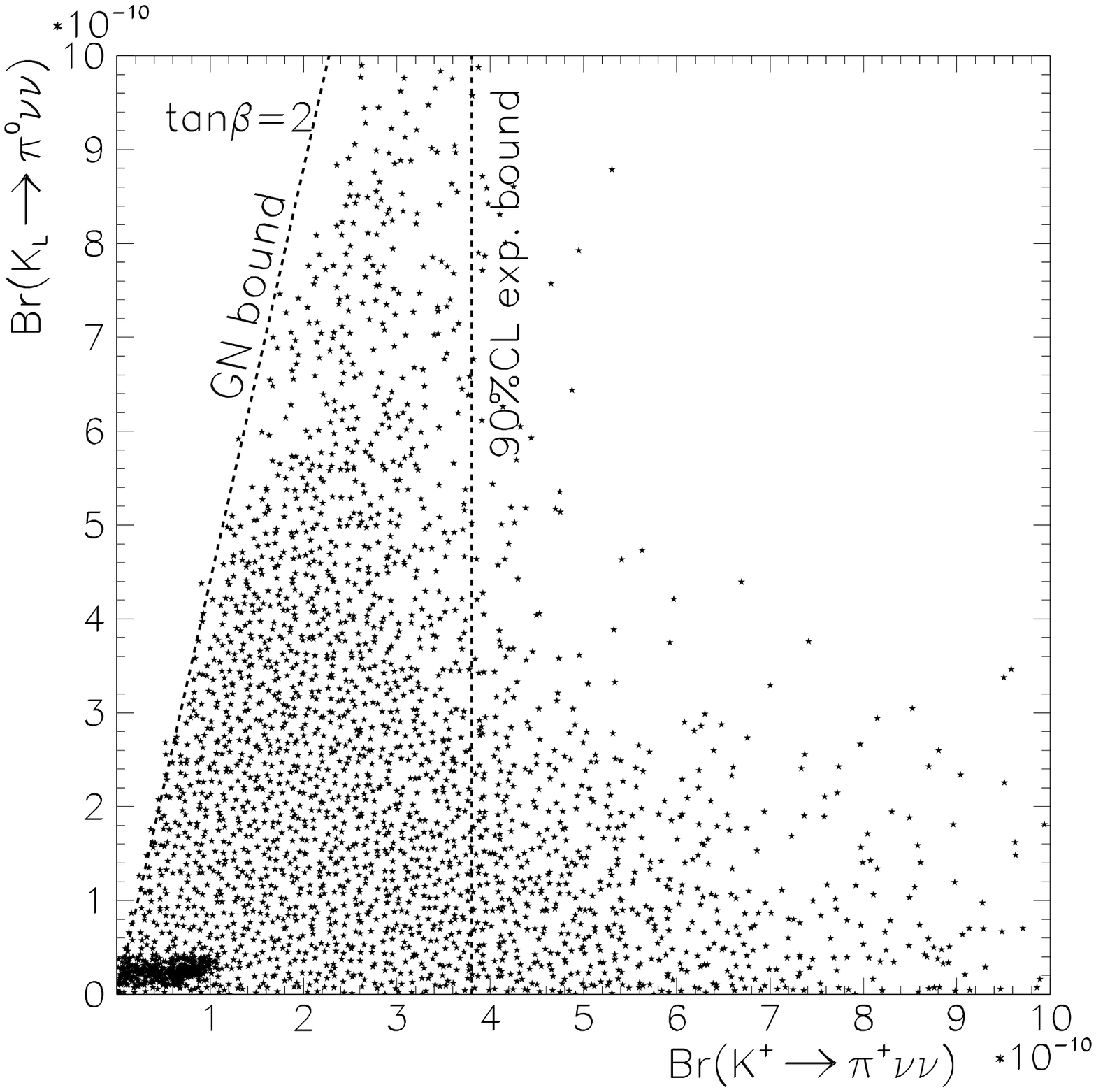,width=0.48\linewidth}
&
\epsfig{file=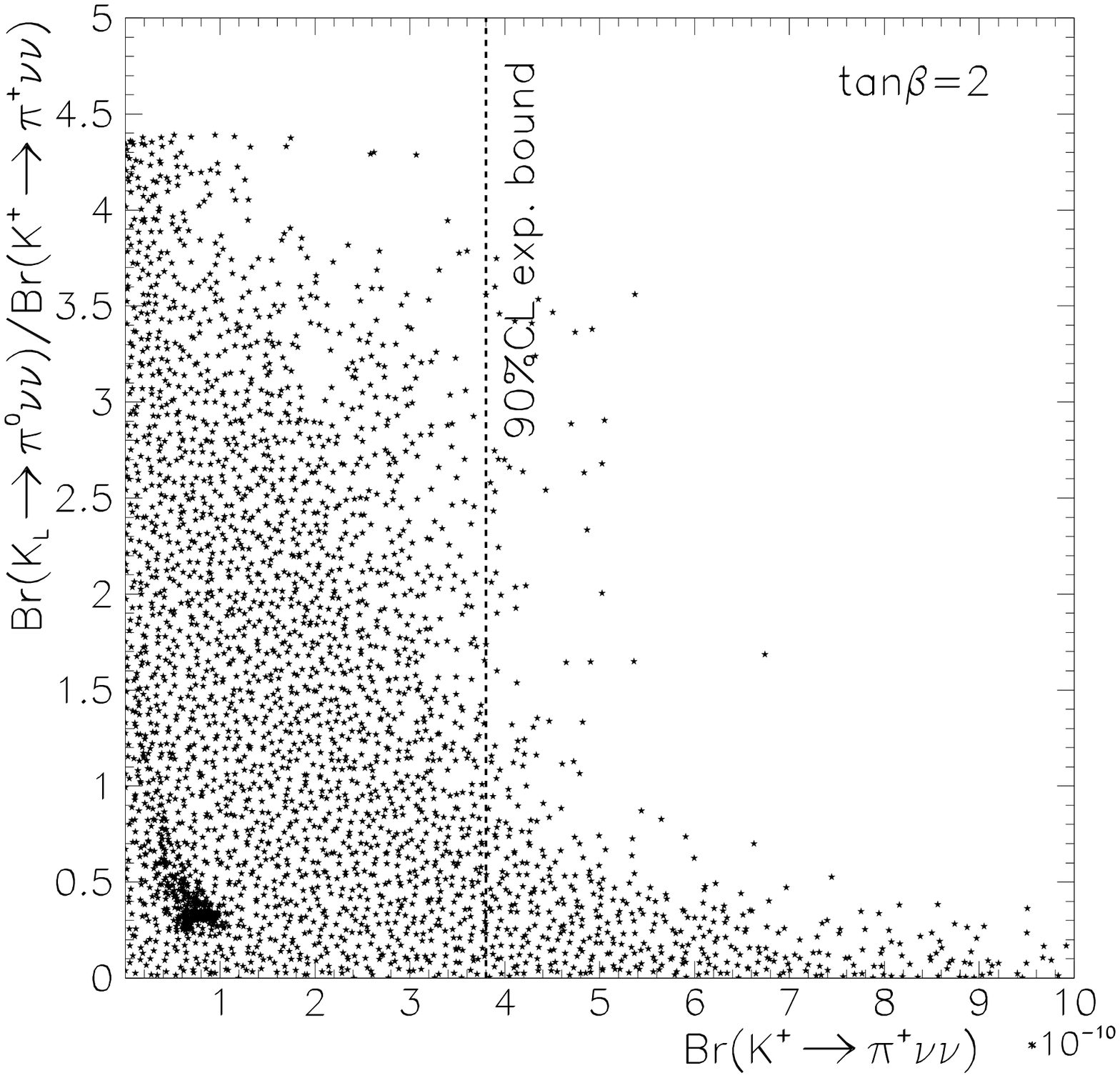,width=0.48\linewidth}
\\
\end{tabular}
\caption{Distributions of $X$, $Br(K_L\ra \pi^0 \nu 
\bar \nu)$ and $Br(K^+\ra \pi^+ \nu \bar \nu)$ for $\tan\beta=2$ in 
the 63-parameter scan.}
\label{fig:fxtb2}
\end{center}
\end{figure}

Plots equivalent to those shown in Fig.~\ref{fig:vegdist} confirm the
assumption that the dependence on most new parameters is weak
(i.e. distributions are flat). Actually, in some cases like left stop
and right sbottom mass parameters they become even flatter than in
Fig.~\ref{fig:vegdist}, as both discussed branching ratios do not
depend directly on them (at least not strongly), and various
experimental constraints can be satisfied varying additional
parameters.

The allowed ranges for $X$ and both branching ratios are extended
somewhat in the 63-parameter scan, but not drastically, as illustrated
in Fig.~\ref{fig:fxtb2}.  In general, as could be expected, it is
easier to generate a $\theta_X$ in the full range, there are more
points with large $Br(\kpn)$, $Br(\klpn)$ or their ratio, but in
general all distributions are still well defined in shape, just a bit
broader.  Thus, the most important conclusion of this kind of
analysis is the statement that even the almost fully general low-energy
MSSM does not lose its predictive power and exploring it can still
lead to reasonable and well defined results.  Although it requires
more effort in numerical computation, it also minimizes the possibility
of overlooking some interesting scenarios which can be realized for
particular SUSY parameter choices.

The numerical analysis described in this section can also be used to put
bounds on the allowed magnitude of the mass insertions in the squark
sector, without resorting to the simplifying assumptions made in the
traditional literature -- absence of cancellations, neglect of interference of
SM and SUSY contributions, the MIA. Results of such an analysis will be presented
elsewhere.

\section{Conclusions}
\label{sec:conclusions}
\setcounter{equation}{0}
In this paper we have analyzed the rare decays $\kpn$ and $\klpn$ in a
general MSSM with conserved R-parity. Our analysis goes beyond those
present in the literature in that
\begin{itemize}
\item
we do not use the MIA but rather work in the
mass eigenstate basis for all particles,
\item
we do not assume the dominance of certain contributions, as done
frequently in the literature, thus allowing for cancellations and
interferences of different diagrams contributing to the decay rates in
question,
\item
we consider a scan of a very large space of parameters, 16 in the
constrained scan and 63 in the extended scan, that to our knowledge has
been presented here for the first time.
\end{itemize}

We should emphasize that the physics scenario considered here differs
from the simple new physics scenario discussed
in~\cite{BFRS-II,BFRS-III} in that the new physics contributions to
$K\to \pi \nu\bar\nu$ in the general MSSM do not only affect the
function $X$ as in \cite{BFRS-II,BFRS-III}, but also modify the values
of the CKM parameters. While the values of $|V_{us}|$, $|V_{ub}|$ and
$|V_{cb}|$, measured in tree-level decays, are the same  in both
papers, the presence of relevant new phases in $B^0_{d,s}-\bar
B^0_{d,s}$ mixings and $\varepsilon_K$ in the general MSSM can
significantly alter the values of $\gamma$, $\beta$ and $R_t$ that are
used in the evaluation of the branching ratios for
$K\to\pi\nu\bar\nu$. Consequently the comparison with the results
presented in \cite{BFRS-III} has to be made with care. For instance
the relation between $\beta_X$ and $\theta_X$ in~(\ref{BX})
can be modified in the
general MSSM through the change of the angle $\beta$. While in
\cite{BFRS-III} one has $\beta=(23.5\pm 2.0)^\circ$, here one
typically finds $12^\circ\le \beta\le 27^\circ$ but this modification
can be compensated by a change of $\theta_X$ so that $\beta_X$
remains unchanged.

More important then is the difference between $R_t$ used
in~\cite{BFRS-III} and $R_t$ found here. As $\gamma=(65\pm7)^\circ$
in~\cite{BFRS-III} but $20^\circ\le\gamma\le 110^\circ$ here (see
Fig.~\ref{fig:vegdist}), the corresponding values of $R_t$ obtained
from~(\ref{VTDG}) with $R_b=0.37$ are $0.86\le R_t\le 0.95$ and
$0.66\le R_t\le 1.18$, respectively.

While these differences in the CKM parameters play some role, they
turn out to be subdominant in comparsison with the differences between
the values for $X$ used here and in~\cite{BFRS-III}:
compare~(\ref{rX}) and Fig.~\ref{fig:fxtb2}.

Concerning individual contributions, in agreement
with~\cite{BRS,COLISI,Buras:1999da} we find that
chargino diagrams are always
strongly dominant, typically one order of magnitude larger than the
neutralino contributions.  Gluino exchanges are fully negligible.
However, unlike these authors we find that in addition to chargino mediated
$Z^0$-penguins, chargino box diagrams can be
important and even dominant for light charged slepton masses.
Thus from the point of view of the
decay rates $Br(K^+\ra \pi^+ \nu \bar \nu)$ and
$Br(K_L\ra \pi^0 \nu \bar \nu)$, 
the general MSSM is a new physics scenario
with enhanced $Z^0$-penguins carrying a new complex phase, similarly
to what has been considered in~\cite{BFRS-III}. However, the presence
of more free parameters than in the simple scenario
in~\cite{BFRS-III}, allows for a broader range of values of $|X|$ and
$\theta_X$ as we stressed above, and there are other manifestations
of new physics such as in modified values of some CKM parameters and in 
box contributions to four-fermion operators.

The answers to the questions posed at the beginning of our paper are
as follows:
\begin{itemize}
\item
The phase $\theta_X$ can be as large as found in
\cite{BFRS-II,BFRS-III}. In fact as seen in Fig.~\ref{fig:xtb} one
finds typically
\be
-160^\circ\le \theta_X\le 50^\circ~
\ee
with a slightly increased range for the extended scan as seen in 
Fig.~\ref{fig:fxtb2}.
Interestingly, there is a visible preference for negative values of
$\theta_X$ that seem to be required by the $B\to\pi K$ data within a
simpler new physics scenario discussed in \cite{BFRS-II,BFRS-III}.
\item
However, among the allowed values of $|X|$, it is not easy to find
simultaneously $|X|\approx 2.2$ and $\theta_X\approx -85^\circ$ as
given in (\ref{rX}). Indeed for $\theta_X\approx -85^\circ$ the
allowed value of $|X|$ in the constrained scan is typically lower than its
SM value of $1.53$ and only in the extended scan can it reach
$|X|\approx 2.0$.  Still, one cannot conclude from this finding that
it is difficult to explain the $B\to\pi K$ data within the general MSSM,
as the relation between the latter decays and $K\to\pi \nu\bar\nu$ in
the general MSSM is rather weak. A very recent study of $B\to\pi K$ decays
in the general MSSM can be found in \cite{Khalil}.
\item
On the other hand for $-50^\circ\le \theta_X\le 50^\circ$, one can
have $|X|$ as high as $7$, implying that in this range of $\theta_X$
very large departures from the SM expectations are possible. A similar
situation is found for $\theta_X=-(150\pm10)^\circ$ with $|X|$
reaching values as high as $7$. This is illustrated in
Figs.~\ref{fig:xtb} and \ref{fig:fxtb2}.
\item
It should be emphasized that values of $|X|$ significantly higher than
$2$, while excluded in the new physics scenario
of~\cite{BFRS-II,BFRS-III} through the data on $B\to X_sl^+l^-$, are
still allowed here because in the general MSSM there are no important
correlations between $B$ and $K$ decays.
\item
The pattern of $Br(\klpn)\approx 3\cdot 10^{-10}$ and $Br(\kpn)\approx
8\cdot 10^{-11}$ found in \cite{BFRS-II,BFRS-III}, although not easy
to achieve within the general MSSM in the constrained scan, as seen in
Figs.~\ref{fig:xtb2} and \ref{fig:xtbp}, can be obtained in the
extended scan as shown in Fig.~\ref{fig:fxtb2}.
\item
More importantly, as seen in Figs.~\ref{fig:xtb2}, \ref{fig:xtbp}
and~\ref{fig:fxtb2}, $Br(\kpn)$ in the ball park of the central
experimental value in (\ref{EXP1}) can be naturally obtained and
simultaneously $Br(\klpn)$ can be even larger than $Br(\kpn)$. An 
example 
of supersymmetric parameters for which such pattern can be obtained 
is shown in table~\ref{tab:example}.
\item
Comparing Figs.~\ref{fig:xtb2} and \ref{fig:xtb} we observe that the 
experimental bound $Br(\kpn)$ in 
(\ref{EXP1}) has a significant impact on the maximal allowed values of $|X|$ 
with
essentially no impact on $\theta_X$. This is not surprising as $\kpn$ is a 
CP-conserving decay. Clearly an improved upper bound on $Br(\klpn)$ should 
have an important impact on the allowed values of $\theta_X$.
\item
Finally, it is interesting to note that even in the general MSSM the angles 
$\beta$ and $\gamma$ of the unitarity triangle are rather constrained,
\be
12^\circ\le\beta\le 27^\circ, \qquad 20^\circ\le \gamma \le 110^\circ~.
\ee
The constraint from the measured asymmetry $a_{\psi K_S}$ in (\ref{APSI}) 
plays an important role here. 
\end{itemize}

In summary, within the general MSSM 
large departures from the SM expectations for $K\to\pi\nu\bar\nu$
are still possible while 
satisfying all existing constraints.
$Br(\kpn)$ and $Br(\klpn)$  can be both 
as large 
as few times $10^{-10}$ with $Br(\klpn)$ often larger than $Br(\kpn)$ and 
close to its model independent upper bound. In particular the results of 
a phenomenological study of enhanced electroweak penguins in 
 \cite{BFRS-III} can be obtained. 
The supersymmetric effects thus turn out to be larger than found 
in \cite{BRS,Buras:1999da}, where typically 
$Br(\kpn)\le 1.7\cdot 10^{-10}$ and $Br(\klpn)\le 1.2\cdot 10^{-10}$ have
been found. 
With regard to~\cite{BRS}, this is partly due to the fact that these authors
use the single MIA, while we use exact mass matrix
diagonalization.
The second reason is the exploration of a much larger space of parameters 
that was  possible in a reasonable time only by using specially designed 
Monte Carlo techniques.
On the other hand, \cite{Buras:1999da} took into account RG effects
between the GUT and weak scales, as is suitable for a high SUSY breaking
scale, which allowed them to constrain the
double mass insertion responsible for the large effects reported
in~\cite{COLISI}. Working in the completely general MSSM, we did not
impose the RG constraint, while our exact mass diagonalization incorporates
the second (and higher) orders in the MIA, in effect reinstating
these large contributions to the extent allowed by the constraints
listed in Subsect.~\ref{subsec:bounds}.

In summary the large enhancements of $Br(\kpn)$ and $Br(\klpn)$ in the 
general MSSM found here are in the spirit of the findings of 
Colangelo and Isidori \cite{COLISI}, except that we effectively incorporate
not only the second but also higher order terms in the MIA, 
we explore much broader range of the space of supersymmetric parameters than 
done by these authors and 
we find the chargino
box contributions more important than found in \cite{COLISI}.

\vspace{5mm}
\noindent
{\bf Acknowledgments}\\
\noindent
We would like to thank Oliver Brein for useful discussions on his
adaptive scanning method, and for reading and comments on
the corresponding parts of our manuscript. S.J. would like to thank
the Fermilab theory group for their hospitality during the final stages
of this work.
T.E. has been supported by the German-Israeli Foundation under the
contract G-698-22.7/2002.
A.J.B. and J.R. have been supported in part by the German
Bundesministerium f\"ur Bildung und Forschung under the contract
05HT4WOA/3 and the DFG Project Bu.\ 706/1-2.
The work of S.J. is supported in part by the DFG
Sonderforschungsbereich/Transregio 9 "Computergest{\"u}tzte Theoretische
Teilchenphysik".
J.R. was supported in part by Alexander von Humboldt Foundation and by
KBN Grant 2 P03B 040 24 (2003-2005).

\newcounter{lp}
\setcounter{lp}{0}
\newcommand{\lpv}{\addtocounter{lp}{1}\noindent\arabic{lp}.~}

\renewcommand{\thesection}{Appendix}

\setcounter{section}{0}
\renewcommand{\thesubsection}{\Alph{section}.\arabic{subsection}}

\setcounter{equation}{0}
\renewcommand{\theequation}{\Alph{section}.\arabic{equation}}

\section{Contributions to the function \boldmath{X}}
\label{app:wilson}

In this appendix we collect the contributions to the functions $X_L$
and $X_R$ within the general MSSM in the mass eigenstates basis for
particles and sparticles. To compactify our notation, we
abbreviate fermion-scalar vertices as $i(V^L P_L + V^R P_R)$, with the
actual values of $V^L$ and $V^R$ given by

\begin{tabular}{ll}
\begin{picture}(140,70)(0,-10)
\DashArrowLine(10,0)(60,0){5}
\Text(0,0)[c]{$D_i^+$}
\ArrowLine(60,0)(110,0)
\Text(120,0)[c]{$N_j$}
\ArrowLine(60,50)(60,0)
\Text(65,45)[l]{$d^I$}
\Vertex(60,0){2}
\end{picture}
&
\raisebox{30\unitlength}{
\begin{minipage}{5cm}
\lefteqn{
V_{dDN}^{LIij} = {\frac{-e}{\sqrt{2}s_Wc_W}} Z^{Ii}_{D}
\left[\frac{1}{3}Z^{1j}_{N} s_W - Z^{2j}_{N} c_W\right] + Y_d^{I} Z_{D}^{(I+3)i}
Z_{N}^{3j}\vspace{1mm}
}
\lefteqn{
V_{dDN}^{RIij} = {\frac{-e\sqrt{2}}{3c_W}} Z_{D}^{(I+3)i} Z_{N}^{1j\star}
 + Y_d^{I} Z_{D}^{Ii} Z_{N}^{3j\star} 
}
\end{minipage}
}
\end{tabular}

\begin{tabular}{ll}
\begin{picture}(140,70)(0,-10)
\DashArrowLine(10,0)(60,0){5}
\Text(0,0)[c]{$U_i^-$}
\ArrowLine(60,0)(110,0)
\Text(120,0)[c]{$C_j^C$}
\ArrowLine(60,50)(60,0)
\Text(65,45)[l]{$d^I$}
\Vertex(60,0){2}
\end{picture}
&
\raisebox{30\unitlength}{
\begin{minipage}{5cm}
\lefteqn{
V_{dUC}^{LIij} = \left[ {\frac{-e}{s_W}} Z_U^{Ji\star} Z_+^{1j} + Y_u^J
Z_U^{(J+3)i\star} Z_+^{2j}\right] V_{JI}\vspace{1mm}
}
\lefteqn{
V_{dUC}^{RIij} = - Y_d^I Z_U^{Ji\star} Z_-^{2j\star} V_{JI} 
}
\end{minipage}
}
\end{tabular}

\begin{tabular}{ll}
\begin{picture}(140,70)(0,-10)
\DashArrowLine(10,0)(60,0){5}
\Text(0,0)[c]{$\tilde{\nu}_K$}
\ArrowLine(60,0)(110,0)
\Text(120,0)[c]{$N_j$}
\ArrowLine(60,50)(60,0)
\Text(65,45)[l]{$\nu^I$}
\Vertex(60,0){2}
\end{picture}
&
\raisebox{30\unitlength}{
\begin{minipage}{5cm}
\lefteqn{
V_{\nu\tilde{\nu}N}^{LIKj} =
{\frac{e}{\sqrt{2}s_Wc_W}} Z^{IK\star}_{\tilde{\nu}}
\left[Z^{1j}_{N} s_W - Z^{2j}_{N} c_W\right]\vspace{1mm}
}
\lefteqn{
V_{\nu\tilde{\nu}N}^{RIKj} = 0
}
\end{minipage}
}
\end{tabular}

\begin{tabular}{ll}
\begin{picture}(140,70)(0,-10)
\DashArrowLine(10,0)(60,0){5}
\Text(0,0)[c]{$L_i^+$}
\ArrowLine(60,0)(110,0)
\Text(120,0)[c]{$C_j$}
\ArrowLine(60,50)(60,0)
\Text(65,45)[l]{$\nu^I$}
\Vertex(60,0){2}
\end{picture}
&
\raisebox{30\unitlength}{
\begin{minipage}{5cm}
\lefteqn{
V_{\nu LC}^{LIij} = {\frac{-e}{s_W}} Z_L^{Ii} Z_-^{1j} - Y_e^I
Z_L^{(I+3)i} Z_-^{2j}\vspace{1mm}
}
\lefteqn{
V_{\nu LC}^{RIij} = 0
}
\end{minipage}
}
\end{tabular}

\noindent where $s_W=\sin\theta_W$, $c_W=\cos\theta_W$ and the definitions of 
the mixing matrices $Z$ and the remaining vertices can be found
in~\cite{ROS}.

The contributions to $X_L$ and $X_R$ can be divided into two classes,
generated by box-type and $Z^0$-penguin diagrams:
\be
  V_{3I}^{} V_{3J}^\star [X_{L,R}]_{JIKL} = \sum_{i=S\!M,h,c,n,g}
\left\{[B^i_{L,R}]_{JIKL}+[P^i_{L,R}]_{JI}\delta_{KL}\right\} .
\label{eq:xgen}
\ee
This formula corresponds to a generalization of~(\ref{Ht}) to
\be
{\tilde{\cal H}}^{(t)}_{\rm eff} = \sum_{K,L=1}^3 V_{3I}^{}
V_{3J}^\star \left[[X_L]_{JIKL} (\bar{d}_Jd_I)_{V-A}
(\bar{\nu}_K\nu_L)_{V-A} + [X_R]_{JIKL}
(\bar{d}_Jd_I)_{V+A}(\bar{\nu}_K\nu_L)_{V-A} \right] ,
\label{eq:hgen}
\ee
where for $K\to\pi\nu\bar{\nu}$ decays one should use $I=1,J=2$ and
$K,L$ are neutrino flavour indices (note that the subscript $L$ on
$X_L$ refers to the chirality structure of the quark current and is no
summation index). In general $X_{L,R}$ quantities defined
in~(\ref{eq:xgen}) carry lepton flavour indices.  Only in the case
when lepton flavour number is conserved in the slepton sector and the
left- and right-slepton flavour diagonal mass parameters are
identical for all three generations, the $X_{L,R}$ are to a good
approximation (neglecting small terms proportional to lepton Yukawa
couplings) universal, as assumed in~(\ref{Ht}). Relation between
$X_{L,R}$ of~(\ref{Ht}) and of~(\ref{eq:hgen}) is given by
\bea
[X_{L,R}]_{21KL} \equiv X_{L,R}\delta_{KL}.
\label{eq:diagx}
\eea
In the numerical analysis in our paper we assume mentioned above
simple flavour conserving structure of the slepton sector and use the
definition~(\ref{eq:diagx}).

Defining further
\be
x_t = \frac{m_t^2}{M_W^2},\qquad y_t = \frac{m_t^2}{M_H^2}
\ee
the non-vanishing box contributions read (summation over all indices
other than $J$, $I$, $K$ and $L$ is understood)
\bea
 [B_L^{SM}]_{JIKL} &=& - V_{3I}^{} V_{3J}^\star 4
 f_1(x_t)\,\delta_{KL}\\[2mm]
 [B_R^h]_{JIKL} &=& - V_{3I}^{} V_{3J}^\star\frac{m_{d_I}m_{d_J}m^2_{e_K}
 \tan^4\beta}{4M_W^2M_H^2}f_1(y_t)\,\delta_{KL}\\[2mm]
 \label{eq:BLc}
 [B_L^c]_{JIKL} &=& - \frac{M_W^2s_W^4}{2e^4} V_{\nu LC}^{LLlm}V_{\nu
 LC}^{LKln\star}V_{dUC}^{LIkm} V_{dUC}^{LJkn\star}\nonumber\\[1mm]
 &\times& m_{C_m} m_{C_n}
 D_0(m^2_{C_m},m^2_{C_n},m^2_{U_k},m^2_{L_l})\\[2mm]
 [B_R^c]_{JIKL} &=& \frac{M_W^2s_W^4}{4e^4} V_{\nu LC}^{LLlm}V_{\nu
 LC}^{LKln\star}V_{dUC}^{RIkm} V_{dUC}^{RJkn\star}
 D_2(m^2_{C_m},m^2_{C_n},m^2_{U_k},m^2_{L_l})\\[2mm]
 [B_L^n]_{JIKL} &=& - \frac{M_W^2s_W^4}{4e^4}V_{dDN}^{LIkm}
 V_{dDN}^{LJkn\star} \bigg[V_{\nu\tilde{\nu}N}^{LLNn}
 V_{\nu\tilde{\nu}N}^{LKNm\star}
 D_2(m^2_{N_m},m^2_{N_n},m^2_{D_k},m^2_{{\tilde \nu}_N})\nonumber\\
 &+& 2V_{\nu\tilde{\nu}N}^{LLNm} V_{\nu\tilde{\nu}N}^{LKNn\star}
 m_{N_m} m_{N_n} D_0(m^2_{N_m},m^2_{N_n},m^2_{D_k},m^2_{{\tilde
 \nu}_N}) \bigg]\\[2mm]
 [B_R^n]_{JIKL} &=& \frac{M_W^2s_W^4}{4e^4}V_{dDN}^{RIkm}
 V_{dDN}^{RJkn\star} \bigg[V_{\nu\tilde{\nu}N}^{LLNm}
 V_{\nu\tilde{\nu}N}^{LKNn\star}
 D_2(m^2_{N_m},m^2_{N_n},m^2_{D_k},m^2_{{\tilde \nu}_N})\nonumber\\
 &+& 2V_{\nu\tilde{\nu}N}^{LLNn} V_{\nu\tilde{\nu}N}^{LKNm\star}
 m_{N_m} m_{N_n} D_0(m^2_{N_m},m^2_{N_n},m^2_{D_k},m^2_{{\tilde
 \nu}_N})\bigg]
\eea
\noindent The non-vanishing penguin contributions read
\bea
 [P_L^{SM}]_{JI} &=& V_{3I}^{}V_{3J}^\star f_2(x_t)\\[2mm]
 [P_L^h]_{JI} &=& -V_{3I}^{}V_{3J}^\star \frac{\cot^2\beta
 M_H^2}{2M_W^2} y_t f_1(y_t)\\[2mm]
 [P_R^h]_{JI} &=& V_{3I}^{}V_{3J}^\star \frac{m_{d_I}m_{d_J}
 \tan^2\beta}{2M_W^2} f_1(y_t)\\[2mm]
 \label{eq:PLc}
 [P_L^c]_{JI} &=& \frac{s_W^2}{8e^2}V_{dUC}^{LIkm} V_{dUC}^{LJin\star}
 \bigg[2 Z_-^{1m} Z_-^{1n\star}\delta_{ki}m_{C_m} m_{C_n}
 C_0(m^2_{U_k},m^2_{C_m},m^2_{C_n})\\[1mm] &-&Z_+^{1n}
 Z_+^{1m\star}\delta_{ki}C_2(m^2_{U_k},m^2_{C_m},m^2_{C_n}) + Z_U^{Nk}
 Z_U^{Ni\star}\delta_{mn} C_2(m^2_{C_m},m^2_{U_k},m^2_{U_i})
 \bigg]\nonumber\\[2mm]
 [P_R^c]_{JI} &=& \frac{s_W^2}{8e^2}V_{dUC}^{RIkm}V_{dUC}^{RJin\star}
 \bigg[2 Z_+^{1n} Z_+^{1m\star}\delta_{ki}m_{C_m} m_{C_n}
 C_0(m^2_{U_k},m^2_{C_m},m^2_{C_n})\\[1mm] &-&Z_-^{1m}
 Z_-^{1n\star}\delta_{ki}C_2(m^2_{U_k},m^2_{C_m},m^2_{C_n}) -
 Z_U^{(N+3)k} Z_U^{(N+3)i\star}\delta_{mn}
 C_2(m^2_{C_m},m^2_{U_k},m^2_{U_i}) \bigg]\nonumber\\[2mm]
 [P_L^n]_{JI} &=& \frac{s_W^2}{8e^2}V_{dDN}^{LIkm} V_{dDN}^{LJin\star}
 \bigg[ Z_D^{(N+3)i} Z_D^{(N+3)k\star}\delta_{mn}
 C_2(m^2_{N_m},m^2_{D_k},m^2_{D_i})\nonumber\\[1mm] &+& (Z_N^{4n}
 Z_N^{4m\star} - Z_N^{3n} Z_N^{3m\star})\delta_{ki}
 C_2(m^2_{D_k},m^2_{N_m},m^2_{N_n})\nonumber\\[1mm] &+&
 2(Z_N^{4n\star} Z_N^{4m} - Z_N^{3n\star} Z_N^{3m})\delta_{ki} m_{N_m}
 m_{N_n} C_0(m^2_{D_k},m^2_{N_m},m^2_{N_n})\bigg]\\[2mm]
 [P_R^n]_{JI} &=& -\frac{s_W^2}{8e^2}V_{dDN}^{RIkm}
 V_{dDN}^{RJin\star} \bigg[Z_D^{Ni} Z_D^{Nk\star}\delta_{mn}
 C_2(m^2_{N_m},m^2_{D_k},m^2_{D_i})\nonumber\\[1mm] &+&(Z_N^{4n\star}
 Z_N^{4m} - Z_N^{3n\star} Z_N^{3m})\delta_{ki}
 C_2(m^2_{D_k},m^2_{N_m},m^2_{N_n})\nonumber\\[1mm] &+&2(Z_N^{4n}
 Z_N^{4m\star} - Z_N^{3n} Z_N^{3m\star})\delta_{ki} m_{N_m} m_{N_n}
 C_0(m^2_{D_k},m^2_{N_m},m^2_{N_n})\bigg]\\[2mm]
 [P_L^{g}]_{JI} &=& \frac{g_s^2 s_W^2}{3e^2}
 Z_D^{(M+3)i}Z_D^{(M+3)k\star} Z_D^{Ik} Z_D^{Ji\star}C_2(m^2_{\tilde
 g},m^2_{D_k},m^2_{D_i})\\[2mm]
 [P_R^{g}]_{JI} &=& -\frac{g_s^2 s_W^2}{3e^2} Z_D^{Mi}Z_D^{Mk\star}
 Z_D^{(I+3)k} Z_D^{(J+3)i\star}C_2(m^2_{\tilde g},m^2_{D_k},m^2_{D_i})
\eea
\noindent The loop functions appearing in these Wilson coefficients
are given by
\bea
f_1(x) &=& \frac{x}{4(1-x)} + \frac{x\ln x}{4(1-x)^2}\\[1mm]
f_2(x) &=& \frac{x(6-x)}{8(1-x)} + \frac{x(2+3x)\ln x}{8(1-x)^2}\\[1mm]
C_0(x,y,z) &=& -{\frac{y}{(x-y)(z-y)}}\ln\frac{y}{x}
 + (y\leftrightarrow z)\label{c0gen}\\[1mm]
C_2(x,y,z) &=& \frac{2}{4 - d} + \log 4\pi - \gamma_E +
 \ln\frac{\mu^2}{x} + 1\nonumber\\[1mm]
 &-& \left\{\frac{y^2}{(x-y)(z-y)}\ln\frac{y}{x}
 + (y\leftrightarrow z)\right\}\label{c2gen}\\
D_0(x,y,z,t) &=& \frac{-y}{(y-x)(y-z)(y-t)}\ln\frac{y}{x}
 + (y\leftrightarrow z) + (y\leftrightarrow t)\label{d0gen}\\[1mm]
D_2(x,y,z,t) &=& \frac{-y^2}{(y-x)(y-z)(y-t)}\ln\frac{y}{x}
 + (y\leftrightarrow z) + (y\leftrightarrow t)\label{d2gen}
\eea
Infinite and $\mu$-dependent terms in $C_2$ always cancels out in
flavour off-diagonal penguins after summation over squark, chargino
and neutralino mixing matrices. We recall that $f_1(x) = B_0(x)$ and
$f_2(x) = C_0(x)$ in the notation of~\cite{BBL}.

\end{document}